\documentclass[twocolumn,prl,superscriptaddress]{revtex4-1}
\usepackage{bm,natbib}
\usepackage{graphicx,color}
\usepackage{mathrsfs}
\usepackage{dcolumn,fancyhdr}
\usepackage{amsmath}
\usepackage{amssymb}
\usepackage{amsfonts,gensymb}
\usepackage{indentfirst,txfonts}
\usepackage{bbold}
\usepackage{multirow}

\newcommand{\ud}{\,\mathrm{d}}

\DeclareMathOperator{\tr}{tr}

\newcommand{\trans}{^{\text{T}}}
\newcommand{\inv}{^{-1}}
\newcommand{\diag}{\text{diag}}

\newlength{\eqboxstorage}

\begin{document}

\title{Experimental determination of irreversible entropy production in out-of-equilibrium mesoscopic quantum systems}
\author{M. Brunelli}%\thanks{These authors contributed equally to this work}
\affiliation{Cavendish Laboratory, University of Cambridge, Cambridge CB3 0HE, United Kingdom}
\author{L. Fusco}%\thanks{These authors contributed equally to this work}
\affiliation{Centre for Theoretical Atomic, Molecular and Optical Physics, School of Mathematics and Physics, Queen's 
University, Belfast BT7 1NN, United Kingdom}
\author{R. Landig}\email{Present address: Department of Physics, Harvard University, Cambridge, Massachusetts 02138, USA.}
\address{Institute for Quantum Electronics, ETH Z\"urich, 8093 Z\"urich, Switzerland}
%\thanks{These authors contributed equally to this work}
\author{W. Wieczorek}
\affiliation{Department of Microtechnology and Nanoscience, Chalmers University of Technology, 412 96 G\"oteborg, Sweden}
\author{J. Hoelscher-Obermaier}
\affiliation{University of Vienna, Faculty of Physics, Vienna Center for Quantum
Science and Technology (VCQ), Boltzmanngasse 5, 1090 Vienna, Austria}
\affiliation{Leibniz University Hannover, %Institut f\"ur Gravitationsphysik,
Institute for Gravitational Physics (Albert-Einstein-Institute), Callinstra\ss e 38, 30167 Hannover, Germany}
\author{G. Landi}
\affiliation{Instituto de F\'isica da Universidade de S\~ao Paulo,  05314-970 S\~ao Paulo, Brazil}
\author{F. L. Semi\~ao}
\affiliation{Centro de Ci\^encias Naturais e Humanas, Universidade Federal do ABC, 09210-170, Santo Andr\'e, S\~ao Paulo, Brazil}
\author{A. Ferraro}
\affiliation{Centre for Theoretical Atomic, Molecular and Optical Physics, School of Mathematics and Physics, Queen's 
University, Belfast BT7 1NN, United Kingdom}
\author{N. Kiesel}
\affiliation{University of Vienna, Faculty of Physics, Vienna Center for Quantum
Science and Technology (VCQ), Boltzmanngasse 5, 1090 Vienna, Austria}
\author{T. Donner}
\affiliation{Institute for Quantum Electronics, ETH Z\"urich, 8093 Z\"urich, Switzerland}
\author{G. De Chiara}
\affiliation{Centre for Theoretical Atomic, Molecular and Optical Physics, School of Mathematics and Physics, Queen's 
University, Belfast BT7 1NN, United Kingdom}
\author{M. Paternostro}
\affiliation{Centre for Theoretical Atomic, Molecular and Optical Physics, School of Mathematics and Physics, Queen's 
University, Belfast BT7 1NN, United Kingdom}

\begin{abstract}
By making use of a recently proposed framework for the inference of 
thermodynamic irreversibility in bosonic quantum systems, we experimentally measure and characterize the entropy production rates 
in the non-equilibrium steady state of two different physical systems -- a micro-mechanical resonator and a Bose-Einstein condensate -- 
each coupled to a high finesse cavity and hence also subject to optical loss. 
Key features of our setups, such as cooling of the mechanical resonator and signatures of a structural quantum phase transition in the 
condensate are reflected in the entropy production rates. Our work demonstrates the possibility to explore irreversibility in driven mesoscopic quantum systems 
and paves the way to a systematic experimental assessment of entropy production beyond the microscopic limit.
\end{abstract}

\maketitle
%\tableofcontents
\date{\today}

%%%%%%%%%%%%%%%%%%%%%%%%%%%%
%
%
%
%
%%%%%%%%%%%%%%%%%%%%%%%%%%%%
%\newpage

\begin{figure}[t!]
\centering
\includegraphics[width=0.8\columnwidth]{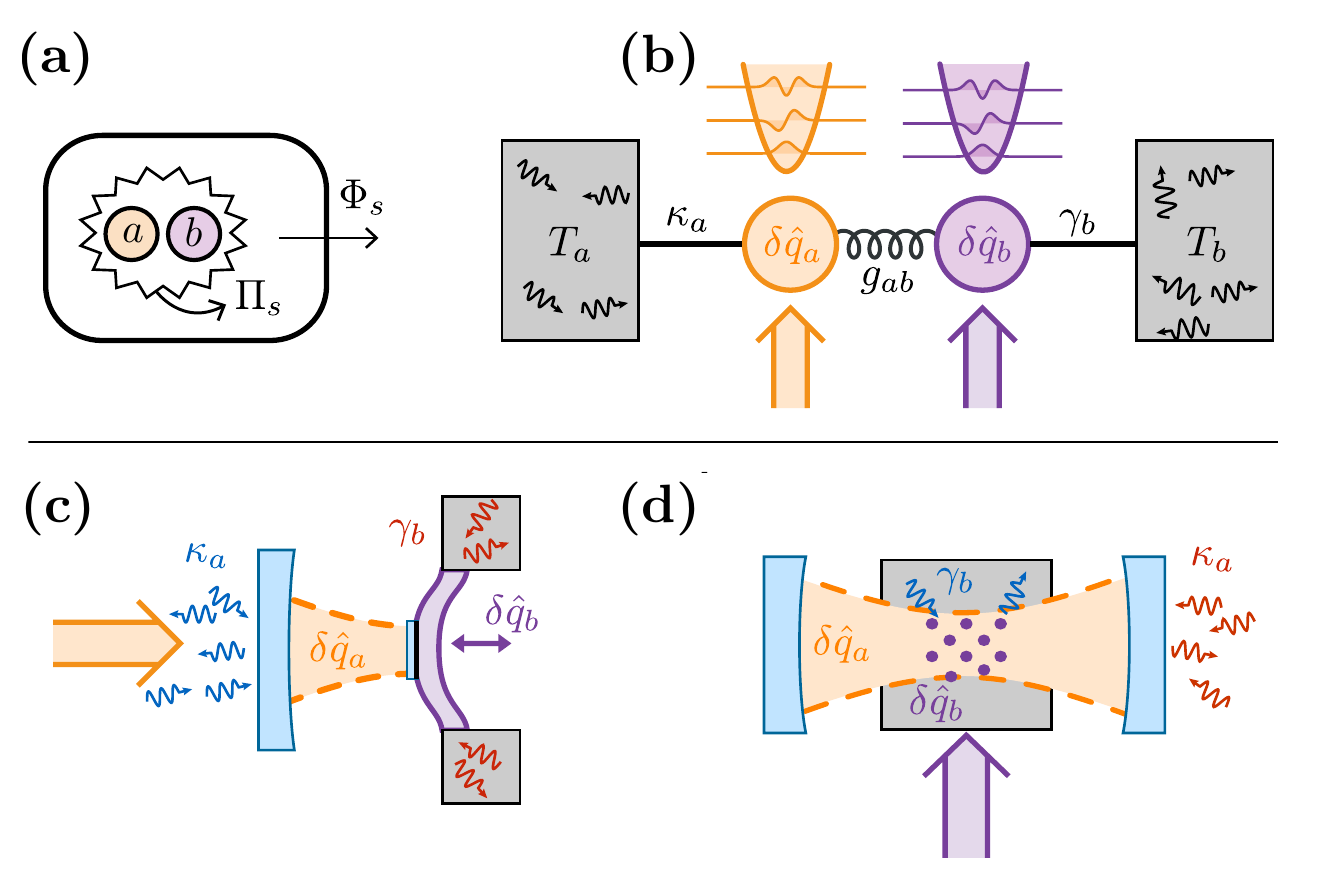} 
\caption{ {\bf (a)} The driven-dissipative system, consisting of the 
coupled subsystems $a$ and $b$, reaches a NESS with an associated entropy production rate $\Pi_s$ and an entropy flux $\Phi_s$ from the system to the 
environment. {\bf (b)} Both systems can be modelled as two quantum harmonic oscillators at frequencies $\omega_a$ and $\omega_b$, linearly coupled with 
a strength $g_{ab}$. Each oscillator is coupled to independent local baths at temperature $T_a$ and $T_b$, respectively. The corresponding coupling rates are 
$\kappa_a$ and $\gamma_b$. The oscillators can be pumped by an external field (purple and orange arrows in the figure).  {\bf (c)} Optomechanical setup: a micro-mechanical oscillator ($\delta\hat{q}_b$) is coupled to the field mode of an optical Fabry-Perot cavity ($\delta\hat{q}_a$). For this setup only the cavity is pumped. {\bf (d)} 
Cavity-BEC setup: the external degree of freedom of a BEC ($\delta\hat{q}_b$) is coupled to the field mode of a cavity ($\delta\hat{q}_a$). For this setup only 
the atoms are pumped. Red and blue wiggly lines indicate heating or cooling of the subsystems via coupling to the baths. In both setups {the number of
excitations in the optical bath is zero, i.e. $n_{T_a}=0$ }.}\label{setup}
\end{figure}

Entropy is a crucial quantity for the characterisation of dynamical processes: it quantifies and links seemingly distant notions such as disorder, information, and 
irreversibility across different disciplinary boundaries~\cite{EntCompl,Zema}. Every finite-time transformation results in some production of entropy,
 which signals the occurrence of irreversibility. Quantifying the amount of irreversible entropy produced by a given process is a goal of paramount importance: {\it entropy production} 
 is a key quantity for the characterisation of non-equilibrium processes, and its minimisation improves the efficiency of thermal machines. The second law of thermodynamics can be formulated in terms of a universal constraint on the {entropy production}, which can never be negative~\cite{StochTherm, EqIneq}. In turn, this leads to the following rate equation for the variation of the entropy 
$S$~\cite{IrrEnt}
\begin{equation}
\frac{dS}{dt}=\Pi (t)-\Phi (t),
\end{equation}
where $\Pi(t)$ and $\Phi(t)$ are the irreversible entropy production rate and the entropy flux from the system to the environment, respectively. When 
the system reaches a non-equilibrium steady-state (NESS) these quantities take values $\Pi_s$ and $\Phi_s$ respectively, such that $\Pi_s = \Phi_s > 0$ 
[see Fig.~\ref{setup} {\bf (a)}]. Under these conditions, entropy is produced and exchanged with the local baths at the same rate. Only when both terms vanish ($\Pi_s=\Phi_s=0$) 
one recovers thermal equilibrium. The entropy production rate directly accounts for the irreversibility of a process and uncovers the non-equilibrium features 
of a system. 
\par
The link between the entropy production rate $\Pi_s$ and irreversibility becomes particularly relevant in small systems subjected to fluctuations, for which a microscopic definition of 
entropy production based on stochastic trajectories of the system has been given~\cite{Seifert}. Experimentally, this notion has been used to test fluctuation 
theorems in a variety of classically operating systems such as a single-electron box~\cite{ExpPekola}, a two-level system driven by a time-dependent potential~
\cite{ExpSeifert}, and a levitated nanoparticle undergoing relaxation~\cite{ExpNovotny}. 
However, in order to harness the working principles of thermodynamic machines working at the quantum level, and pinpoint the differences between their performances and those of their classical counterparts, it is important to analyse the entropy generated through genuine quantum dynamics~\cite{arrow}. %Such knowledge would be key in optimising the performance of such machines. 
Moreover, while so far nanoscale systems have been used for the experimental study of classical out-of-equilibrium thermodynamics, irreversible entropy production arising from quantum dynamics in mesoscopic quantum systems has not been experimentally investigated yet.

%Recently, progress towards the theoretical characterisation of entropy production in bosonic systems brought out of equilibrium has been made~\cite{Landi,Santos,Matteo}. In this paper, we make use of such theoretical framework to quantify experimentally the amount of irreversibility in the NESS of a quantum system by coupling it to a probe embodied by a light field mode, that is used for detection purposes. Therefore, our study addresses the estimation of $\Pi_s$ for two coupled quantum systems (one being the probe itself) and quantifies it in terms of relevant, controllable parameters. As such, our results show how a key indicator of irreversibility is fully within the grasp of dynamically controlled quantum dynamics. 
%%In its generality, irreversible entropy production is worthwhile to be investigated in a variety of different experimental contexts. In particular, 

Recently, progress towards the theoretical characterisation of entropy production in bosonic systems brought out of equilibrium has been made~\cite{Landi,Santos,Matteo}. In this paper, we make use of such theoretical framework to quantify experimentally the amount of irreversibility in the NESS of two different driven-dissipative quantum systems, realized by coupling bosonic systems to high-finesse cavities. The light field mode of a cavity allows to infer the entropy production in terms of relevant controllable parameters of the coupled system. 
In particular, in this study, we investigate the influence of different dynamical regimes and sources of environmental noise on the quantum fluctuations of a quantum system, and thus the corresponding entropy production rate.  In order to address such influences, we assess two distinct experimental setups: a cavity-optomechanical (cavity-OM) device and a Bose-Einstein condensate (BEC) with cavity-mediated long-range interactions~\cite{Aspelmeyer,Baumann,mottl}. The required measurements are based on the spectra of the light fields leaking out of the respective cavities.
Remarkably, the entropy production reflects the specific features of the two experimental platforms, which are very different in nature despite the common description provided here. As such, our results show how a key indicator of irreversibility is fully within the grasp of dynamically controlled quantum dynamics.

%When addressing non-equilibrium quantum processes, it is interesting to investigate how different dynamical regimes and sources of environmental noise influence the quantum fluctuations of a quantum system, and thus the corresponding entropy production rate. In order to address such influences, we assess two distinct experimental setups: a cavity-optomechanical device and a BEC with cavity-mediated long-range interactions~\cite{Aspelmeyer,Baumann,mottl}. The required measurements are based on the spectra of the light fields leaking out of the respective cavities. Remarkably, the entropy production reflects the specific features  of the two experimental platforms addressed in our study, which are very different in nature despite the common description provided here.
\par

In cavity-OM systems, the cavity photon number is coupled to the position of the mechanical oscillator [cf. Fig.~\ref{setup}{\bf (b)} and {\bf (c)}]. Our specific 
implementation uses a Fabry-Perot cavity. One of its mirrors is a doubly clamped, highly reflective, mechanical cantilever. Radiation pressure 
couples the intra-cavity photon number to the position of the cantilever. The mechanical support of the cantilever provides a local heat bath at room temperature. 
The optical cavity is driven by a laser that is red-detuned by the mechanical frequency from the optical cavity resonance. For a driving 
laser without classical noise, the cavity mode is coupled to a zero-excitation heat bath. We observe sideband cooling of the mechanical motion~\cite{Schliesser, 
Groeblacher, Chan, Teufel} and, for large drive powers, strong optomechanical coupling
~\cite{Groeblacher2,Verhagen,Teufel2}. To analyse 
the entropy production rate of the cavity-OM system, we measure the light reflected off the cavity via homodyne detection.

Also in the second implementation, the two coupled harmonic oscillators correspond to a light field mode coupled to a mechanical degree of freedom [cf. Fig~\ref{setup}{\bf (b)} and {\bf (d)}]. We load a BEC into a high-finesse optical cavity and illuminate the atoms with a standing-wave transverse laser field. Far-off resonant 
scattering of photons from the laser field into a near-detuned, initially empty cavity field mode, couples the zero-momentum mode of the BEC to an excited momentum 
mode. The scattering process mediates effective atom-atom interactions, which are of long-range, since the photons are delocalized in the cavity mode~\cite{mottl}. This interaction 
is tunable in strength via the power of the transverse laser beam. The long-range interaction can be brought to competition with the kinetic energy of the atoms, resulting 
in a structural phase transition~\cite{ZurichNew}. In the spatially homogeneous phase, for increasing interaction, the energy of the excited momentum mode softens, until at a critical 
interaction strength the system breaks a discrete symmetry and the atoms arrange in a spatially modulated density distribution. The equivalence of this system to a 
Dicke model has been demonstrated in Ref.~\cite{Baumann}. We measure the cavity light field leaking through the mirrors with a heterodyne detection setup. 
The spectral analysis of this signal is used to infer the diverging amount of atomic density fluctuations accompanying the structural phase transition~\cite{ZurichNew}.

%\begin{widetext}
\begin{table}[t!]
\begin{center}
\begin{tabular}{cccccccc}
\hline
\hline
& ${\omega_a/2\pi}$ & $\kappa_a/2\pi$ & $\omega_b/2\pi$ & $\gamma_b/2\pi$ & $T_b$  & $\text{Other}$\\ [0.5ex]
& [MHz] & [kHz] & [kHz] & [Hz] & [K] & \text{parameters}\\ [0.5ex]
\hline
cavity-OM &$1.27815$&435.849  & $1278.15 $  & 264.1 & 292 & $m=176$ng \\ 
\hline
cavity-BEC&$15.13 $&1250   & 8.3  & \cite{SI} &$38\times 10^{-9}$ & $N=10^5$\\ 
\hline
\hline
\end{tabular}
\end{center}
\caption{{Physical parameters for the two experimental setups}. 
The damping rate $\gamma_b$ 
is constant in the cavity-OM experiment, while in the cavity-BEC setup it 
depends on the actual working point (cf. Ref.~\cite{SI} for details). Here, $m$ is the effective mass of the mechanical oscillator, and $N$ is the number of $^{87}$Rb atoms in the BEC.}
\label{table:tab1}
\end{table}
%\end{widetext}

\begin{figure}[b]
\centering
\includegraphics[width=\columnwidth]{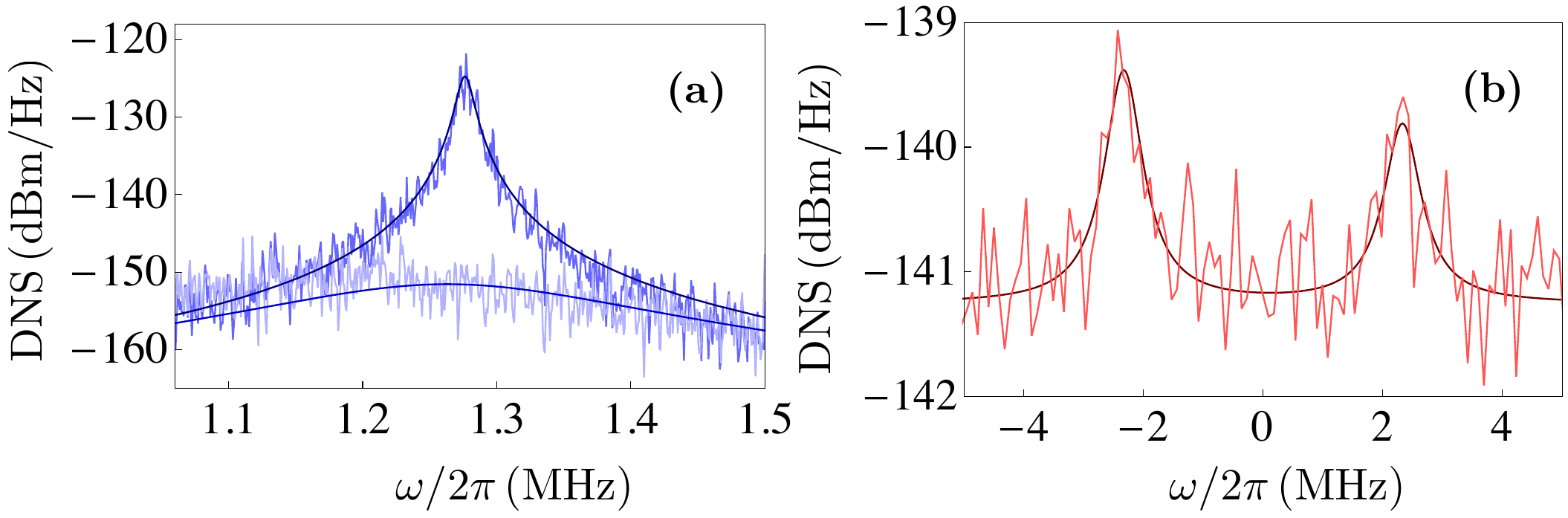}%\includegraphics[width=0.5\columnwidth]{DNSJoint16Dic(b).pdf}
\caption{ Experimental density noise spectra. { Panel {\bf (a)}: Density noise spectrum (DNS) of the phase quadrature of the output cavity field, attenuated before detection, for the cavity-OM setup}. The jagged blue curve refers to a value of the rescaled coupling
$g_{ab}/\kappa_a=0.49$, while the jagged light-blue curve to $g_{ab}/\kappa_a=2.29$. The fits of the DNS are shown as smooth lines. Notice that the power spectrum is originally dimensionless, and has been here converted to SI units for uniformity of notation.  
Panel {\bf (b)}: DNS of the extra-cavity field for the cavity-BEC system at a coupling $(g_{ab}/g_{ab}^\mathrm{cr})^2 = 0.93$. A fit of the DNS is shown as a smooth line.}
\label{DNS}
\end{figure}

In both cases, the effective interaction between the oscillators is obtained by a harmonic expansion of the field operators around their mean values, resulting in two linearly 
coupled quantum oscillators [cf. Fig.~\ref{setup}{\bf (b)}]. We denote with $\delta \hat{q}_{a,b}$ and $\delta \hat{p}_{a,b}$ the position and momentum fluctuation operators around the mean-field values 
of the two oscillators. In what follows, $a$ and $b$ refer to the optical and mechanical/atomic oscillators, respectively. In a frame rotating at the frequency 
$\omega_p$ of the respective pump fields, the oscillators have frequencies $\omega_a=\omega_c-\omega_p$ and $\omega_b$ (here $\omega_c$ is the frequency of 
the cavity field). Their interaction is described by the Hamiltonian
\begin{equation}\label{HamiltonianPaper}
\hat{H}=\frac{\hbar \omega_a}{2} (\delta\hat{q}_a^{2}+\delta\hat{p}_a^{2})+\frac{\hbar \omega_b}{2} (\delta\hat{q}_b^{2}+\delta\hat{p}_b^{2})+\hbar g_{ab}\delta\hat{q}_a
\delta\hat{q}_b,
\end{equation}
where $g_{ab}$ is the coupling strength between the modes. In the superradiant phase of the Dicke model, an additional squeezing term of the atomic mode must be included 
in the Hamiltonian~\cite{mottl}. For the derivation of the models and the values of the parameters in the two setups, we refer to Refs.~\cite{SI} and to Table~
{\ref{table:tab1}}. The systems are inherently open: each harmonic oscillator is independently coupled to a local bath. This provides both a dissipation channel and extra 
quantum fluctuations in addition to those present in the closed systems. The optical cavity mode is coupled to the surrounding electromagnetic vacuum with a decay rate $\kappa_a$. 
On the other hand, the nature of the mechanical/atomic bath is specific to the setup being considered. In the cavity-OM system, the coupling of the vibrating mirror to 
the background of phonon modes is described in terms of quantum Brownian motion. In the cavity-BEC system, dissipation is due to the collection of excited 
Bogolioubov modes, which provides a bath for the condensate. In both cases, we assume  oscillator $b$ to be in contact with a Markovian bath at temperature $T_b$ and 
rate $\gamma_b$. The average number of excitations in the equilibrium state of oscillator $b$ is thus $n_{T_b}=(e^{\hbar \omega_b/ k_B T_b}-1)^{-1}$ (cf. Ref~\cite{footnote}). 
The driven-dissipative nature of the systems is such that a NESS is eventually reached~\cite{dickeETH,Aspelmeyer}.

The linear dynamics undergone by the coupled oscillators allows us to exploit a framework developed for linear stochastic processes~\cite{Landi,Santos,Matteo}. In particular, the situation that we face is perfectly suited to the use of the framework for the quantification of entropy production proposed in Ref.~\cite{Santos}, where the entropy $S$ of an arbitrary bosonic quantum system prepared in a Gaussian state is written in terms of the Shannon entropy of the Wigner function
\begin{equation}
S(t)=-\int {\cal W}(u,t) \log {\cal W}(u,t)du,
\end{equation}
where ${\cal W}(u,t)$ is the Wigner function at time $t$ corresponding to the state of the two oscillators, and $u$ is the corresponding vector of complex phase-space variables. The quadratic nature of Eq.~\eqref{HamiltonianPaper} and the initial thermal state of the oscillators in both setups ensures the positivity of ${\cal W}(u,t)$ and allows us to write it in terms of the variances of the fluctuation operators of the oscillators, which enormously simplifies the explicit calculation of $\Pi(t)$. In the NESS, all entropy produced in the system flows to the environments so that $\Pi_s = \Phi_s$. Following the lines sketched in Ref.~\cite{SI}, the entropy production rate in the NESS due to the quantum fluctuations takes the form
\begin{equation}\label{Pi_ss}
\Pi_{s} =\Phi_s=2\gamma_b\left( \frac{ n_b+1/2}{n_{T_b}+1/2} - 1 \right) + 4\kappa_a n_a=\mu_b+\mu_a,
\end{equation}
where $n_a=\langle(\delta\hat{q}_a^{2}+\delta\hat{p}_a^{2}-1)\rangle_s/2$ and $n_b=\langle(\delta\hat{q}_b^{2}+\delta\hat{p}_b^{2}-1)\rangle_s/2$ are the average number of 
excitations in the NESS of the two oscillators in excess of the zero-point motion of the respective harmonic oscillator. 
In the cavity-OM expression for $\mu_b$, instead of the full phonon number $n_b$, only the momentum variance $\langle\delta\hat p_b^2\rangle_s$ enters as we assume Brownian motion damping. 

Eq.~\eqref{Pi_ss} represents our main theoretical result: it quantifies the entropic contribution, ascribable to the quantum fluctuations that the system has to pay to remain in its 
NESS. It thus directly quantifies the irreversibility of the driven-dissipative dynamics of two linearly coupled quantum oscillators, well beyond the linear-response limit. For vanishing coupling the systems reach thermal equilibrium (i.e. $n_a=0$ and $n_b=n_{T_b}$), and $\Pi_s$ vanishes. Moreover, there is no dependence on the correlations between the oscillators, since in a NESS the entropy 
production rate $\Pi_s$ equals the flux rate $\Phi_s$. Thus, the entropy flux from the system to the overall environment determines the amount of {\it irreversibility} produced 
within the driven-dissipative model, and is directly linked to the breaking down of the microscopic detailed balance~\cite{arrow}. The previous considerations also allow us to identify two contributions to $\Pi_s$, linked to the mechanical/atomic and optical oscillator, referred to as $\mu_{a}$ and 
$\mu_b$, respectively. They are the individual entropy flows to each environment and show how the entropy produced in the NESS is split into two distinct fluxes. We note that the explicit form of Eq.~\eqref{Pi_ss} in terms of the sum of such independent terms strongly relies on the local nature of the environments that we have cconsidered, and we expect it not to hold in more general situations. %are not entropy production rates for subsystem $j=a,b$, inasmuch as they do not meet the requirements of positivity. This is made apparent by looking at the mechanical contribution to $\Pi_s$ in Eq.~\eqref{Pi_ss}, which can take negative values.
  The dissipative evolution arising from the contact 
with the environments is manifested explicitly in Eq.~\eqref{Pi_ss} by the presence of the rates $\gamma_b$ and $\kappa_a$. In both settings, the mechanical/atomic damping 
rate $\gamma_b$ is much smaller than the cavity decay rate $\kappa_a$, as can be appreciated from Table~\ref{table:tab1}. 
\par
\begin{figure}[t!]
\centering
\includegraphics[width=\columnwidth]{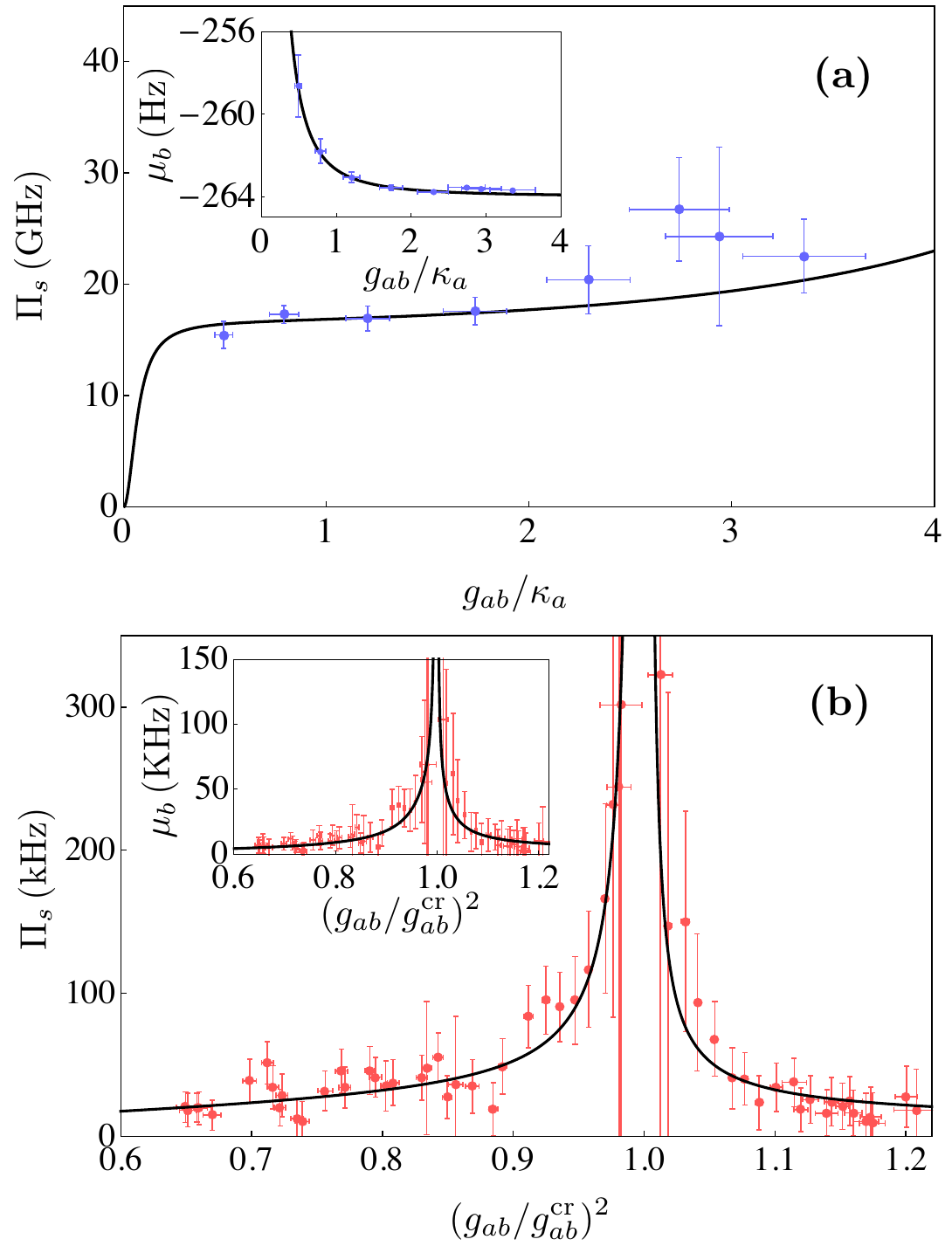}
\caption{ Experimental assessment of the irreversible entropy production rate $\Pi_s$ at the NESS for {\bf (a)}  the cavity-OM system and {\bf (b)} the cavity BEC system. In the cavity-OM system, $g_{ab}$ is twice the standard optomechanical coupling rate~\cite{Aspelmeyer,SI}. For the cavity-BEC 
setup, the control parameter $g_{ab}$ is renormalised with respect to the critical parameter $g_{ab}^{\text{cr}}=\sqrt{(\kappa_a^2+\omega_a^2)\omega_b/4\omega_a}$. The insets 
show the behaviour of $\mu_b$ in each of the settings considered. In both panels, the solid black lines show the theoretical predictions based on the values given in Table I. The blue and red dots show the 
experimental data for the cavity-OM and cavity-BEC experiment, respectively. In panel {\bf (a)}, the vertical error bars report statistical errors extracted from the fit, while the
horizontal ones show experimental error on the values of the parameter. In panel {\bf (b)}, the vertical and horizontal error bars report the statistical errors from the fit and the 
determination of the critical point, respectively~\cite{ZurichNew}. }
\label{EntroSS}
\end{figure}

A general formulation of entropy production demands the knowledge of the global state of the system~\cite{REF,REF1,REF2,REF3,REF4}. However, $\Pi_s$ evaluated for the linearised dynamics in Eq.~\eqref{HamiltonianPaper} only involves the mean excitations of the oscillators~\cite{Landi,Matteo}. For the experimental regime of interest, the dynamics of the cavity field adiabatically follows the mechanical/atomic mode. By measuring the light field leaking out of the cavity we thus can infer about both $\mu_a$ and $\mu_b$. For both experimental setups, the coupling $g_{ab}$ is 
varied by increasing the power of the pump. The density noise spectrum (DNS) of the cavity field quadratures is recorded~\cite{ZurichNew,Wieczorek}. 
Typical examples of the experimental DNS, together with the fitting curves used for their analysis, are shown in Fig.~\ref{DNS}. In the cavity-OM experiment, the datasets 
are taken for $\omega_a=\omega_b$, which is the working point where the cooling of the mechanical resonator is most effective in the resolved-sideband regime. In the cavity-BEC experiment, on the other 
hand, the parameters are $\omega_a\gg \omega_b$, resulting in only a tiny admixture of the optical subsystem. A further difference between the two platforms is in the way the 
two oscillators are populated: in the optomechanical case, we have $n_b \gg n_a$ for the lowest coupling values, while they become comparable in size for the maximum 
cooling achieved. In the cavity-BEC setup, the cavity field is considerably less 
populated than the atomic mode. Finally, the mechanical bath is at room temperature, while the temperature of the atomic reservoir is below the condensation 
point and in the nK range (cf. Table~\ref{table:tab1}). This highlights and reinforces the diversity of the experimental 
platforms that we have addressed within a unique framework for the quantification of irreversible entropy. 

Following the technical approach illustrated in Refs.~\cite{Landi,Santos,Matteo} and sketched in \cite{SI}, we have separately reconstructed the two terms $\mu_a$ and $\mu_b$ that determine quantitatively $\Pi_s$. Fig.~\ref{EntroSS} displays the experimental data together with the theoretical model, demonstrating a very good quantitative agreement.
Besides the influences of the environments, the entropy production rates depends on the interplay between the \emph{mutual dynamics} of the 
oscillators. For the cavity-OM system, the contribution to $\Pi_s$ we observe from the mechanical oscillator is much smaller than the one coming from the optical field. On the contrary, $\mu_a\simeq\mu_b$ in the atomic setup. For each of the two experiments $\Pi_s$ is positive, in agreement with the second law. In the cavity-OM setup, $\mu_{a}$ is an increasing 
function of the coupling: the stronger the pump, the further the system operates away from thermal equilibrium and the more entropy is generated. At the same time, 
$\mu_{b}$ takes negative values, whose magnitude increases for increasing values of $g_{ab}$. This is legitimate as $\mu_b$ is not {\it per se} an entropy production rate but represents an individual flux, which can thus take negative values (while $\mu_a+\mu_b$ has to be positive). 
The observed behaviour of $\mu_{b}$ is a signature of optomechanical cooling: its growth, in absolute value, with $g_{ab}$ shows the increase of the entropy flow from the mechanical resonator 
to the cavity field, corresponding to lowering of the effective temperature of the resonator. As for the cavity-BEC system, the divergent behaviour of the entropy production rate 
at the critical point reflects the occurrence of the structural phase transition: at $g_{ab}^{\text{cr}}$, the known divergence of the populations of the two oscillators at the steady-state~\cite{domokos} results in the singularity of both $\mu_a$ and $\mu_b$ separately. The irreversible entropy production rate thus diverges at criticality. 

We have experimentally determined the entropy production rate, a key indicator of irreversibility, in driven-dissipative quantum 
systems operating at the steady-state. The two experimental setups, being instances of mesoscopic systems undergoing quantum dynamics, allowed us to link the phenomenology of the entropy production rate to the salient features of their physics. We have thus assessed architectures that could embody the building blocks of a generation of future thermodynamic machines working out of equilibrium, and thus subjected to irreversible processes. For such devices, the quantification of irreversibility will be very relevant for the characterisation of their efficiency, as it will provide useful information to design protocols able to quench it, thus optimising their working principles.

{\it Acknowledgements.--} We are grateful to M. Aspelmeyer, T. Esslinger, J. Goold, I. Lesanovsky, E. Lutz, and J. Schmiedmayer for useful comments and 
fruitful discussions during the development of this project. We thank S. Gr\"oblacher for support with microfabrication, and F. Brennecke and R. Mottl for support in taking and evaluating the cavity-BEC data. This work was supported by the European Union through the projects TherMiQ, TEQ, SIQS, iQOEMS, and ITN cQOM, the European Research Council through the Advanced Grant project SQMBS, the Brazilian CNPq through Grant number 302900/2017-9 and the ``Ci\^{e}ncia sem Fronteiras'' programme via the ``Pesquisador Visitante 
Especial'' initiative (Grant number 401265/2012-9), the S\~ao Paulo Research Foundation (FAPESP) under grant number 2014/01218-2, the Brazilian National Institute of Science and Technology of Quantum 
Information (INCT/IQ), the Vienna Science and Technology Fund (WWTF, Project No. ICT12-049), and the Austrian Science Fund FWF (JHO: W1210, CoQuS; NK: AY0095221, START).

%\end{document}

\newpage
\onecolumngrid
\renewcommand{\bibnumfmt}[1]{[S#1]}
\renewcommand{\citenumfont}[1]{S#1}
\section*{Supplementary Information}
\vspace{1cm}
\twocolumngrid
\title{Supplementary Informations on\\"Experimental determination of irreversible entropy production in out-of-equilibrium mesoscopic quantum systems"}
\author{M. Brunelli}%\thanks{These authors contributed equally to this work}
\affiliation{Cavendish Laboratory, University of Cambridge, Cambridge CB3 0HE, United Kingdom}
\author{L. Fusco}%\thanks{These authors contributed equally to this work}
\affiliation{Centre for Theoretical Atomic, Molecular and Optical Physics, School of Mathematics and Physics, Queen's 
University, Belfast BT7 1NN, United Kingdom}
\author{R. Landig}
\affiliation{Department of Physics, Harvard University, Cambridge, Massachusetts 02138, USA}
\author{W. Wieczorek}
\affiliation{Department of Microtechnology and Nanoscience, Chalmers University of Technology, 412 96 G\"oteborg, Sweden}
\author{J. Hoelscher-Obermaier}
\affiliation{University of Vienna, Faculty of Physics, Vienna Center for Quantum
Science and Technology (VCQ), Boltzmanngasse 5, 1090 Vienna, Austria}
\affiliation{Leibniz University Hannover, %Institut f\"ur Gravitationsphysik,
Institute for Gravitational Physics (Albert-Einstein-Institute), Callinstra\ss e 38, 30167 Hannover, Germany}
\author{G. Landi}
\affiliation{Instituto de F\'isica da Universidade de S\~ao Paulo,  05314-970 S\~ao Paulo, Brazil}
\author{F. L. Semi\~ao}
\affiliation{Centro de Ci\^encias Naturais e Humanas, Universidade Federal do ABC, 09210-170, Santo Andr\'e, S\~ao Paulo, Brazil}
\author{A. Ferraro}
\affiliation{Centre for Theoretical Atomic, Molecular and Optical Physics, School of Mathematics and Physics, Queen's 
University, Belfast BT7 1NN, United Kingdom}
\author{N. Kiesel}
\affiliation{University of Vienna, Faculty of Physics, Vienna Center for Quantum
Science and Technology (VCQ), Boltzmanngasse 5, 1090 Vienna, Austria}
\author{T. Donner}
\affiliation{Institute for Quantum Electronics, ETH Z\"urich, 8093 Z\"urich, Switzerland}
\author{G. De Chiara}
\affiliation{Centre for Theoretical Atomic, Molecular and Optical Physics, School of Mathematics and Physics, Queen's 
University, Belfast BT7 1NN, United Kingdom}
\author{M. Paternostro}
\affiliation{Centre for Theoretical Atomic, Molecular and Optical Physics, School of Mathematics and Physics, Queen's 
University, Belfast BT7 1NN, United Kingdom}

\maketitle
%\tableofcontents
\date{\today}

\newpage
\onecolumngrid
\renewcommand{\bibnumfmt}[1]{[S#1]}
\renewcommand{\citenumfont}[1]{S#1}
%\section*{Supplementary Information}
\vspace{1cm}
\twocolumngrid
%\section{Methods}
\section{Experimental methods for the cavity-OM setting} 
The mechanical oscillator is a doubly-clamped Si$_3$N$_4$ cantilever with a resonance frequency of the fundamental, 
out-of-plane mode at 1.278MHz. It couples, with the damping rate $\gamma_b$, to its mechanical support, which is at a temperature of 292K. A distributed Bragg reflector 
microfabricated from a Ta$_2$O$_5$/SiO$_2$ stacked on top of the cantilever maximizes its reflectivity ($> 99.995 \% $). This allows coupling the mechanical oscillator via radiation 
pressure to the optical cavity field. The $10$mm long optical cavity has a finesse of 17200 and is formed by an input mirror with a reflectivity of $99.97\%$ and the cantilever. It is 
operated in high vacuum ($10^{-6}$mbar), such that coupling to the background gas is negligible. The optical cavity is driven by two laser fields with orthogonal polarization, the 
driving and auxiliary laser beam, respectively. The laser fields are derived from the same source with a wavelength of 1064\,nm. The driving laser field is red-detuned with respect 
to a cavity mode by one mechanical frequency. Its optomechanical interaction results in sideband cooling of the mechanical motion down to 0.45K. For drive powers larger than 
5.4\,mW we observe strong coupling $g_{ab}\gg2\kappa_a$ between the optical cavity and the mechanical oscillator. Here $\kappa_a$ is the decay rate of the electromagnetic field in the optical cavity. For the present experiment, we change $g_{ab}$ between 
$0.5\kappa_a$ and $3.3\kappa_a$ by varying the optical power sent to the cavity from $0.3$mW to $15$mW. The auxiliary laser field is kept resonant with the optical cavity and has 
a constant optomechanical coupling of $0.4\kappa_a$ throughout the experiment. It allows us to keep the detuning of the driving laser field stable. The driving field is measured via 
homodyne detection in reflection off the optical cavity. We choose the measured optical quadrature by setting the relative phase between the local oscillator and the signal beam. 
For further details on the experimental implementation see Ref.~\cite{SWieczorek}.

\section{Experimental methods for the cavity-BEC setting} 
A BEC of $1.0(1) \times 10^5$ atoms ($^{87}$Rb) is prepared at the location of the mode of the ultra-high finesse optical cavity. The atoms are transversally illuminated by a standing wave laser field at a wavelength of $785.3$nm. The power of the transverse laser is linearly increased over $500$ms, thereby crossing the critical point of the phase transition. Light leaking out of the resonator is directed to a heterodyne detection system. The resulting electronic signal of the balanced photodiodes is mixed down to a frequency of $50$kHz, amplified and filtered before being digitalized with an analog-to digital converter for further analysis.

The temperature of the atoms after preparation is 20(10) nK, determined from absorption images. During the experiment, the temperature of the atoms increases to 38(10) nK at the critical pump power due to off-resonant scattering processes. At the same time, the atom number decreases due to trap losses by 26\%. This loss is accounted for in the analysis by the according scaling of the relative coupling strength.

To evaluate the data, we divide the time signal of the two demodulated heterodyne quadratures, which we label $Q_1$ and $Q_2$, into half overlapping subtraces of $11$ms length. For each subtrace, we calculate the Fourier spectrum of $Q_1 + iQ_2$ and subsequently convert it into a density noise spectrum [cf. Fig.~2 {\bf (b)} of the main paper]. For further details on the experimental implementation and the data evaluation see Ref.~\cite{SZurichNew}.

\section{Entropy production at the steady-state}
In what follows, we report details on the theoretical treatment and the experimental measurements for the optomechanical and atomic setups.

The dynamics of the fluctuations around the semi-classical steady-state can be described in terms of Langevin equations 
for the dimensionless quadrature operators $\delta\hat{q}_b=\bigl(\delta\hat{b}+ \delta\hat{b}^{\dagger})/\sqrt2$,  $\delta \hat{q}_a=\bigl(\delta\hat{a}+ \delta\hat{a}^{\dagger})/\sqrt2$ and their conjugate 
momenta $\delta \hat{p}_b$, $\delta \hat{p}_a$, where $a$ and $b$ refer to the optical and mechanical/atomic oscillators, respectively. They can be arranged in the vector $u(t) = (\delta \hat{q}_b, \delta \hat{p}_b, \delta \hat{q}_a, \delta \hat{p}_a)^\mathrm{T}$, with $\langle u(t) \rangle\equiv0$. The effect of 
the local baths then enters in the form a vector of the input noise operators $N(t)=(\sqrt{2\gamma_b}\hat{q}_b^{\text{in}},\sqrt{2\gamma_b}\hat{p}_b^{\text{in}},\sqrt{2\kappa_a}\hat{q}_a^{\text{in}},
\sqrt{2\kappa_a}\hat{p}_a^{\text{in}})^\mathrm{T}$. In the optomechanical case quantum Brownian motion only couples to the mechanical momentum, and hence the first entry of $N(t)$ is zero. %, i.e., $[N(t)]_{1}=0$.\\

The linear dynamics implies that the Wigner distribution ${\cal W}(u,t)$ of the two harmonic oscillators is a positive Gaussian function in the 
quantum phase space, whose evolution is described by Fokker-Planck equations, and hence a complete description of the system can be given in terms of the second statistical moments of the fluctuation operators. 
These can be arranged in the covariance matrix $\sigma$, defined as $\sigma_{ij}(t):=\langle\{u_i(t),u_j(t)\}\rangle/2$. As the system reaches the stationary state the covariance matrix 
$\sigma_s=\lim_{t\rightarrow \infty} \sigma(t)$ satisfies the equation $A \sigma_s + \sigma_s A\trans = - D$,  where $A$ and $D$ are referred to as drift matrix and diffusion matrix, respectively; their explicit
expression will be provided for both the systems.
\par
For non linear systems the entropy production is written in terms of integrals of probability currents, an analytical expression of which is in general impossible to provide \cite{SREF}. In what follows we
show instead that the entropy production rate assumes a very simple expression
in terms of the elements of the covariance matrix. 
\par
The total rate of change of the entropy of the global system is
\begin{equation}\label{entropy_balance}
\frac{\ud S}{\ud t} = \Pi(t) - \Phi(t),
\end{equation}
where $\Pi(t)$ is the entropy production rate of the system and $\Phi(t)$ is the entropy flux rate, from the system to the environment. 

Following the framework set in Ref.~\cite{Santos} for the quantification of the entropy production in a Gaussian bosonic system undergoing a non-equilibrium quantum process, we use the  Shannon entropy of the Wigner function to calculate the entropy $S$ of the system. We thus have
\begin{equation}
S(t)=-\int {\cal W}(u,t) \log {\cal W}(u,t)du,
\end{equation}
where
\begin{equation}
{\cal W}(u,t)=\frac{1}{\pi^2\sqrt{|\sigma|}}\exp\left\{-\frac{1}{2}u^T\sigma^{-1}u\right\},
\end{equation}
$|\sigma|$ being the determinant of the covariance matrix $\sigma$. It has been shown that this is a suitable quantifier of the informational content of a Gaussian state \cite{SAdesso}.
In order to address irreversibility in the dynamics, it is necessary to distinguish between even and odd variables under time reversal operation, which is equivalent to inverting the sign of momentum. At the covariance-matrix level, this operation is implemented by the matrix $E=\diag(1,-1,1,-1)$.
A given function of the dynamical variables can be decomposed as
\begin{equation}
f(u,t)=f^{\text{irr}}(u,t)+f^{\text{rev}}(u,t),
\end{equation}
where the irreversible part is even under time reversal $f^{\text{irr}}(u,t)=Ef^{\text{irr}}(Eu,t)$, while the reversible part $f^{\text{rev}}(u,t)=-Ef^{\text{rev}}(Eu,t)$ is odd.  
When applying this decomposition to the drift matrix, $A=A^{\text{irr}}+A^{\text{rev}}$, we have
\begin{equation}\label{Airr}
A^{\text{irr}}=\text{diag}\left(-\gamma_b,-\gamma_b,-\kappa_a,-\kappa_a\right) \, ,
\end{equation}
and
\begin{equation}
A^{\text{rev}} =
\begin{pmatrix}
0 & \omega_b& 0 & 0 \\
-\omega_b& 0& g_{ab} & 0 \\
0 & 0& 0 & \omega_a \\
g_{ab}& 0& -\omega_a & 0 \\
\end{pmatrix},
\end{equation}
where, as mentioned in the main text, $g_{ab}$ is the coupling between the two oscillators, $\omega_b$ is the frequency of the atomic/mechanical oscillator, and $\omega_a$ is the detuning between the cavity mode frequency and the pump frequency. These parameters are introduced in the main text with the Hamiltonian describing the closed dynamics of the two coupled oscillators
\begin{equation}\label{Hamiltonian}
\hat{H}=\frac{\hbar \omega_a}{2} (\delta\hat{q}_a^{2}+\delta\hat{p}_a^{2})+\frac{\hbar \omega_b}{2} (\delta\hat{q}_b^{2}+\delta\hat{p}_b^{2})+\hbar g_{ab}\delta\hat{q}_a
\delta\hat{q}_b.
\end{equation}

In Refs.~\cite{SLandi,SMatteo} it was shown that the entropy production rate can be cast in the following form  
\begin{equation}\label{Pi}
\Pi(t) = \frac{\ud S}{\ud t} + \tr(2 A^{\text{irr}\,T}  D\inv A^{\text{irr}} \sigma + A^{\text{irr}})\, ,
\end{equation}
where the pseudo-inverse of $D$ is
\begin{equation}\label{Dinv}
D^{-1}=\text{diag}\left(\frac{1}{\gamma_b(2n_{T_b}+1)},\frac{1}{\gamma_b(2n_{T_b}+1)},\frac{1}{\kappa_a},\frac{1}{\kappa_a}\right) \, .
\end{equation}
Again, in the optomechanical case we have $[A^{\text{irr}}]_{11}=0$ and $[D^{-1}]_{11}=0$.\\
The entropy flux rate $\Phi(t)$ can be read directly from Eqs.~(\ref{entropy_balance}) and (\ref{Pi}) as being simply the second term in the RHS of Eq.~(\ref{Pi}). We can
parametrize $\sigma$ as
\begin{equation}\label{sigma}
\sigma = \begin{pmatrix}
\bf{M} & \bf{C} \\
\bf{C}\trans & \bf{F} 
\end{pmatrix}
=
 \begin{pmatrix}
m_1 & m_{12} & c_{11} & c_{12} \\
m_{12} & m_2 & c_{21} & c_{22} \\
c_{11} & c_{21} & f_{1} & f_{12} \\
c_{12} & c_{22} & f_{12} & f_2
\end{pmatrix},
\end{equation}
where ${\bf M}$ and ${\bf F}$ describe the properties of the mechanical/atomic and field modes respectively, while ${\bf C}$ denotes the correlations.
We can thus write the entropy production in Eq.~(\ref{Pi}) as 
\begin{equation}
\begin{aligned}
\Pi(t) = \frac{\ud S}{\ud t}&+ \gamma_b\left( \frac{m_1(t)+m_2(t)}{n_{T_b}+1/2} - 2 \right)+ 2\kappa_a (f_1(t)+f_2(t)-1),
\end{aligned}
\end{equation}
which in terms of the fluctuation number operators is
\begin{equation}\label{Pi_nice}
\Pi(t) = \frac{\ud S}{\ud t} +2\gamma_b\left( \frac{\langle \delta\hat{b}^{\dagger}\delta\hat{b}\rangle+1/2}{n_{T_b}+1/2} - 1 \right) + 4\kappa_a \langle \delta\hat{a}^{\dagger}\delta\hat{a}\rangle.
\end{equation}
At the steady-state it becomes
\begin{equation}\label{Pi_ss}
\Pi_{s} = 2\gamma_b\left( \frac{\langle \delta\hat{b}^{\dagger}\delta\hat{b}\rangle_{s}+1/2}{n_{T_b}+1/2} - 1 \right) + 4\kappa_a \langle \delta\hat{a}^{\dagger}\delta\hat{a}\rangle_{s}.
\end{equation}
The very same calculation in the optomechanical case, given the different properties of the bath, leads to the slightly different, but conceptually similar expression
\begin{equation}\label{Pi_ss}
\Pi_{s} = \gamma_b\left( \frac{\langle \delta\hat{p}^2\rangle_{s}}{n_{T_b}+1/2} - 1 \right) + 4\kappa_a \langle \delta\hat{a}^{\dagger}\delta\hat{a}\rangle_{s}.
\end{equation}

\section{Dynamics of the cavity-OM system}
In this Section the details regarding the cavity-OM system are discussed, including the description of its dynamics and of the detection scheme. 
For the sake of clarity, we will adopt the notation commonly used in the field, and then make the explicit correspondence to the more abstract notation employed 
in the main text, common to both the systems.
The system of interest consists of a Fabry-Perot cavity with a vibrating end mirror. We denote with $\hat{a}$ $([\hat{a},\hat{a}^{\dagger}]=1)$ the 
annihilation operator relative to the cavity field oscillating at frequency $\omega_c$, and with $\hat{x}=\bigl(\hat{a}+\hat{a}^{\dagger}\bigr)/\sqrt 2$ 
and $\hat{y}=i\bigl( \hat{a}^{\dagger}-\hat{a}\bigr)/\sqrt 2$ the quadratures of the field. Analogously $\hat{b}$ $([\hat{b},\hat{b}^{\dagger}]=1)$
describes a quantized mode of the mechanical resonator relative to  frequency $\omega_m$, and $\hat{q}=\bigl(\hat{b}+\hat{b}^{\dagger}\bigr)/\sqrt 2$ and $\hat{p}=i\bigl( \hat{b}^{\dagger}-\hat{b}\bigr)/\sqrt 2$ its dimensionless position and momentum variables.
The strength of the radiation pressure interaction is quantified by the single-photon coupling rate $g_0=\frac{\omega_c}{(L/x_\mathrm{zpf})}$, where  
$L$ is the cavity length and $x_\mathrm{zpf}=\sqrt{\hbar/ m \omega_\mathrm{m}}$ is the zero-point fluctuation term of the mechanical position in the 
ground state, being $m$ the effective mass of the mechanical resonator. The cavity is also driven by a laser field oscillating at frequency $\omega_p$, 
which couples to the cavity through the fixed end-mirror. In a frame rotating at the frequency of the pump the Hamiltonian of the two modes coupled 
by radiation-pressure interaction reads
  \begin{equation}\label{Hamiltonian}
\hat{H}=\hbar\widetilde{\Delta}\hat{a}^{\dagger}\hat{a}+\frac{\hbar \omega_m}{2} \bigl(\hat{q}^2+\hat{p}^2\bigr) - \hbar g_0 \hat{a}^{\dagger}\hat{a} \hat{q}
+i \hbar\mathcal{E}(\hat{a}^{\dagger}-\hat{a}) \, ,
\end{equation}
where $\widetilde{\Delta}=\omega_c-\omega_p$ and we set $\mathcal{E}=\sqrt{2 P \kappa_1/ \hbar \omega_p}$, being $P$ the incident laser power and 
$\kappa_1$ the input-coupler decay rate. The dynamics of the system is also affected by the presence of the environment. 
Specifically, the mechanical mode is in contact with a bath at finite temperature, and then affected by a viscous force with damping rate $\gamma_m$ 
and by a Brownian stochastic force with zero mean value $\hat{\xi}(t)$, satisfying the correlation function~\cite{Sgiovannetti}
\begin{equation}
\langle \hat{\xi}(t)\hat{\xi}(t')\rangle = \frac{\gamma_m}{\omega_m}\int_{-\infty}^{\infty} \frac{d \omega}{2 \pi} e^{- i \omega(t-t')} \omega \left[ \coth 
\left(\frac{\hbar \omega}{2 k_B T} \right)+1\right] \, .
\end{equation}
Then, due to the nonzero transmission of the cavity mirrors, the 
cavity field is also affected by losses, modeled by quantum noise input operators obeying the following correlation functions
\begin{equation}
\langle \hat{a}_j^{in}(t)\hat{a}_j^{in, \dagger}(t')\rangle= \delta (t-t')\, ,
\end{equation}   
with $j=1,2$, the other correlation functions being zero. Note that in order to properly model the experimental setup, we distinguish between two different loss mechanisms, 
and hence two different loss channels $\hat{a}_1^{in}$ and $\hat{a}_2^{in}$:  $\kappa_1$ is the decay rate of the coupling port (fixed mirror) while $\kappa_2$ 
is a term collecting the internal losses of the cavity, due to unwanted effects like absorption at the mirror or scattering processes into spurious modes. Such contributions are additive, and we thus employ the total cavity decay rate $\kappa=\kappa_1+\kappa_2$ when needed. 
\par
Since the cavity is driven by an intense field,  provided that the system remains in a stable regime, a steady configuration will be reached, characterized by
a displaced position of the mirror and a new intra-cavity amplitude $\alpha_s$. In a mean-field spirit, one can assume small fluctuations around this classical steady-state, 
and by standard linearization obtain the following set of Langevin equations for the quantum fluctuation operators~\cite{SAspelmeyer}
\begin{align}\label{LinLang}
\delta \dot{\hat{q}}&=\omega_m \delta \hat{p} \, , \nonumber \\
\delta \dot{\hat{p}}&=-\omega_m \delta \hat{q} -\gamma_m \delta \hat{p} + G \delta \hat{x} + \hat{\xi} , \\
\delta \dot{\hat{x}}&=-\kappa \delta \hat{x}+ \Delta \delta \hat{y} +\sqrt{2 \kappa_1} \hat{x}_1^{in}+\sqrt{2 \kappa_2} \hat{x}_2^{in} ,\nonumber \\
\delta \dot{\hat{y}}&=-\kappa \delta \hat{y}- \Delta \delta \hat{x} + G \delta \hat{q} +\sqrt{2 \kappa_1} \hat{y}_1^{in}+\sqrt{2 \kappa_2} \hat{y}_2^{in}, \nonumber  
\end{align}
where we have introduced the quadratures operators $\hat{x}_j^{in}=(\hat{a}_j^{in}+\hat{a}_j^{in, \dagger})/\sqrt{2}$ and 
$\hat{y}_j^{in}=i(\hat{a}_j^{in, \dagger}-\hat{a}_j^{in})/\sqrt{2}$. In the set of equations (\ref{LinLang}) the enhanced optomechanical coupling  is given by
$G=\frac{\sqrt{2}g_0\vert \mathcal{E}\vert}{\sqrt{\kappa^2+\Delta^2}}$ with $\Delta=\widetilde{\Delta}-\frac{g_0^2\vert \alpha_s\vert^2}{\omega_m}$.
The former equations can be arranged in a more compact form as a matrix equation for the vector of the fluctuations 
$\delta \hat{u}=(\delta \hat{q},\delta \hat{p},\delta \hat{x},\delta \hat{y})^T$: 
\begin{equation}\label{LinLangVec}
\delta \dot{\hat{u}}(t)=A \delta \hat{u}(t) + \hat{N}(t) \, ,
\end{equation} 
where the drift matrix $A$  is given by 
\begin{equation}\label{A}
A=
\left(
\begin{array}{cccc}
 0 & \omega_m & 0 & 0 \\
 -\omega_m & -\gamma_m  & G & 0 \\
 0 & 0 & -\kappa  & \Delta \\
 G & 0 & -\Delta & -\kappa 
\end{array}
\right) \, ,
\end{equation}
while $\hat{N}=(0,\hat{\xi},\sqrt{2\kappa_1}\hat{x}^{in}_1+\sqrt{2\kappa_2}\hat{x}^{in}_2,\sqrt{2\kappa_1}\hat{y}^{in}_1+\sqrt{2\kappa_2}\hat{y}^{in}_2)^T$
is the vector containing the noise operators. 
\par
If we move to the frequency domain $\delta \hat{u}_j(t)=\int \frac{d \omega}{2 \pi} e^{- i \omega t} \delta \hat{u}_j(\omega)$, the system of linear differential equations 
(\ref{LinLangVec}) becomes a set of algebraic equations and can thus be easily solved, once endowed with the proper set of correlation functions 
\begin{align}
\langle \hat{x}_i^{in}(\omega)\hat{x}_j^{in}(\omega')\rangle&= \langle \hat{y}_i^{in}(\omega)\hat{y}_j^{in}(\omega')\rangle= \pi \delta (\omega+\omega')\delta_{ij} \, , \\ 
\langle \hat{x}_i^{in}(\omega)\hat{y}_j^{in}(\omega')\rangle&= \langle \hat{y}_i^{in}(\omega)\hat{x}_j^{in}(\omega')\rangle^*= i \pi \delta (\omega+\omega')\delta_{ij} \, , \\
\langle \hat{\xi}(\omega)\hat{\xi}(\omega')\rangle&=2 \pi \frac{\gamma_m}{\omega_m} \omega  \left[ \coth 
\left(\frac{\hbar \omega}{2 k_B T} \right)+1\right]\delta (\omega+\omega') \, .
\end{align}
Formally the solution, as a function of the quantum noise operators, is given by
\begin{equation}
\delta{\hat{u}}(\omega)=i (\omega-i A)^{-1} \hat{N}(\omega) \, .
\end{equation}
It is now easy to draw the correspondence between the notation used in the main text, common to both the cavity-OM and cavity-BEC setups, and
the notation employed in this section. We summarise it in the following table, and we notice in particular that $g_{ab}$ is twice the standard optomechanical 
coupling $G$.
\begin{center}
\begin{tabular}{ |c||c|c|c| } 
\hline
 & Cavity-OM notation & Our common notation \\
\hline
\hline
\multirow{3}{5em}{Mechanical mode} & $\hat{q},\, \hat{p}$ & $\hat{q}_b,\, \hat{p}_b$ \\ 
& $\omega_m$ & $\omega_b$ \\ 
& $\gamma_m$ & $\gamma_b$ \\ 
\hline
\multirow{3}{4em}{Cavity field} & $\hat{x},\, \hat{y}$ & $\hat{q}_a,\, \hat{p}_a$ \\ 
& $\Delta$ & $\omega_a$ \\ 
& $\kappa$ & $\kappa_a$ \\ 
\hline
\multirow{1}{4em}{Coupling} & $2G$ & $g_{ab}$ \\ 
\hline

\end{tabular}
\end{center}

\subsection{Detection of the extra-cavity signal}  
We can then compute the symmetrized two-point correlation function in the frequency space both for the amplitude and phase quadrature, which can
be cast in the following form
\begin{equation}
\left\langle \left\{ \delta \hat{x}(\omega),\delta \hat{x}(\omega')\right\}\right\rangle/2=2\pi\delta(\omega+\omega') S_x(\omega)\, , 
\end{equation}
and
\begin{equation}
\left\langle \left\{ \delta \hat{y}(\omega),\delta \hat{y}(\omega')\right\}\right\rangle/2=2\pi\delta(\omega+\omega') S_y(\omega) \, ,
\end{equation}
where the delta distributions account for the stationarity of the process and $S_x$ and $S_y$, referred to as the density noise spectra of the field quadratures, are 
given by
\begin{align}\label{S_x}
S_x(\omega)&=\frac{1}{\vert d(\omega)\vert^2}\left\{ \kappa (\Delta^2+\kappa^2+\omega^2) \left\vert \omega(\omega+i \gamma_m)-\omega_m^2 \right\vert^2 \right.  \nonumber \\
& + \left. G^2\Delta^2 \omega_m \gamma_m \omega  \coth \left(\frac{\hbar \omega}{2 k_B T} \right)   \right\}
\end{align}
and
\begin{align}\label{S_y}
S_y(\omega)&=\,\frac{1}{\vert d(\omega)\vert^2}\left\{ \kappa \left\vert \Delta\left[\omega(\omega+i \gamma_m)-\omega_m^2\right]+\Delta G^2\omega_m \right\vert^2 \right.  \nonumber \\
& + \left. \kappa (\kappa^2 + G^2) \left\vert \omega(\omega+i \gamma_m)-\omega_m^2 \right\vert^2 \right. \nonumber \\
& \left. + \,G^2(\kappa^2 + G^2) \omega_m \gamma_m \omega  \coth \left(\frac{\hbar \omega}{2 k_B T} \right)   \right\} \, ,
\end{align}
where $d(\omega)=\left[\Delta^2+(\kappa_1+\kappa_2-i \omega)^2\right]\left[\omega(\omega+i \gamma_m)-\omega_m^2\right]+\Delta G^2\omega_m$, and
in the range of parameters of our interest it is safe to consider the Markovian limit of the phononic bath. 
On the other hand, if we solved the linearized Langevin equations expressing the intra-cavity quadrature $\delta \hat{Y}$ in terms of the mirror quadrature fluctuation 
$\delta \hat{q}$, we would find that the intra-cavity adiabatically follows the mirror position. The information about the dynamics of the mechanical mode is then imprinted 
in the phase of the cavity field. This feature directly reflects in the behavior of the noise spectrum, namely~\cite{Sclerk}   
\begin{equation}\label{S_y,S_q}
S_y(\omega)= \frac{G^2(\kappa^2+\omega^2)}{\vert \Delta^2+(\kappa-i \omega)^2\vert^2} \frac{S_q(\omega)}{2 \pi}+
\frac{\kappa(\kappa^2+\omega^2+\Delta^2)}{\vert \Delta^2+(\kappa-i \omega)^2\vert^2} \, ,
\end{equation}
where $S_q(\omega)$ is the spectral density of the mechanical position and is given by
\begin{align}\label{S_q}
\frac{S_q(\omega)}{2\pi}&=  \,\frac{\omega_m^2}{\vert d(\omega)\vert^2}\left[(\kappa^2+\omega^2+\Delta^2)^2-4\Delta^2 \omega^2\right] \nonumber \\
& \times \, \left\{\frac{\kappa G^2(\kappa^2+\omega^2+\Delta^2)}
{\left[ \kappa^2+(\Delta-\omega)^2\right]\left[ \kappa^2+(\Delta+\omega)^2\right]} \right. \nonumber \\
&\left. + \,\frac{\gamma_m}{\omega_m} \omega  
\coth \left(\frac{\hbar \omega}{2 k_B T} \right) \right\} \, .
\end{align}
For the parameters used in the experiment (see later) the second term in the right-hand side of Eq. (\ref{S_y,S_q}) can be neglected, so that $S_y(\omega)$ and 
$S_q(\omega)$ are related in a very simple way. The importance of Eq. (\ref{S_y,S_q}) is that it will enable us to infer the relevant dynamical features of the mechanical mode --- 
included its contribution to the entropy production rate --- from the optical spectrum. In particular we can access the oscillator variances, and thus its average
 energy, by integrating the spectral density in the frequency domain, namely   
\begin{equation}\label{MechEnergy}
\langle \delta \hat{q}^2\rangle=\int_{-\infty}^{\infty} \frac{d \omega}{2\pi} S_{q}(\omega)\, , \qquad  
\langle \delta \hat{p}^2\rangle=\int_{-\infty}^{\infty} \frac{d \omega}{2\pi} \frac{\omega^2}{\omega_m^2}S_{q}(\omega) \, . 
\end{equation}
\par
Finally, since the detection necessarily takes place outside the cavity, we need to move from intra-cavity to extra-cavity variables. This can be done employing 
the following generalized input-output relations
\begin{equation}
\hat{a}_1^{out}(t)=\sqrt{\eta}\bigl[ \sqrt{2\kappa_1}\delta \hat{a}(t) - \hat{a}_1^{in}(t)\bigr] + \sqrt{1-\eta} \, \hat{c}_1^{out}(t) \, .
\end{equation}
A few comments are in order. Firstly, only mode $\hat{a}_1$ is involved in moving extra-cavity, since mode $\hat{a}_2$ is associated with irreversible losses. 
Secondly, we need to take into account a finite efficiency in the detection process, that comes from modeling an imperfect detector as a perfect one preceded by a 
beam splitter of transmissivity $\eta \in [0,1]$, which mixes the output signal with an uncorrelated field $\hat{c}_1^{out}$. Computing the two-frequency auto-correlation 
function of the quadrature $\hat{y}^{out}_1$, we obtain
the following expression for the density noise spectrum 
\begin{equation}\label{S_yout}
S_{y_1}^{out}(\omega)= \eta \left[2 \kappa_1 \frac{S_y(\omega)}{2 \pi} -\sqrt{2 \kappa_1} \Re\mathrm{e} [\beta(\omega)] +\frac12\right] + (1-\eta)
\end{equation}
where $\beta(\omega)=d(\omega)^{-1} \sqrt{2\kappa} (\kappa- i\omega)[\omega(\omega+i \gamma_m)-\omega_m^2]$ and 
$\beta(\omega)^*=\beta(-\omega)$. We can also recognize an additive contribution due to the shot noise . An analogous expression can be derived for 
$S_{x_1}^{out}$. Again, by integrating the latter in the frequency space we access the second moments of the quadratures, and then the average energy
\begin{equation}\label{energySy}
\begin{aligned}
\langle (\hat{x}^{out}_1)^2\rangle&=\int_{-\infty}^{\infty} \frac{d \omega}{2\pi} S_{x_1}^{out}(\omega)\, , \qquad  \\
\langle (\hat{y}^{out}_1)^2\rangle&=\int_{-\infty}^{\infty} \frac{d \omega}{2\pi} S_{y_1}^{out}(\omega) \, . \\
\end{aligned}
\end{equation}

\section{Dynamics of the Cavity-BEC system: the Dicke model}
In this section we discuss the details of the theoretical treatment and experimental investigation of the cavity-BEC system. Again, as presented in the previous section for the cavity-OM experiment, we will use the notation commonly used in the cavity-BEC literature, and then make the correspondence to the notation common to both systems.\\
The physical system of interest is a Bose-Einstein Condensate (BEC) of $N$ atoms inside an ultrahigh-finesse optical cavity. The atoms are pumped transversally with a far-detuned standing-wave laser field \cite{SdickeETH}.
It has been shown that the Hamiltonian of this system, for the closed case, maps to the Dicke model, which will be described in the following sections \cite{SBaumann}. The mapping is based on the fact that the far-detuned laser field pumping the atoms makes possible to adiabatically eliminate the internal excited states of the atoms. The states having a role in the dynamics of the system are thus given by the eigenstates of momentum realising the external degree of freedom. A solution of the dynamics, in a mean-field description, of the system shows that the significantly populated momentum states are  only two: the homogeneous ground state, labelled as $\psi_0$, and the superposition of the first excited momentum states in both the cavity and transverse pump axes labelled as $\psi_1$. The states $\psi_0$ and $\psi_1$ differ in energy by twice the recoil energy $\hbar \omega_0:=\hbar^2 k^2/m$, where $k$ is the wavector of both the cavity field mode and the laser field mode, and $m$ is the atomic mass. Physically this is due to the transverse pump field that couples the excited momentum mode $\psi_1$ of the BEC to the cavity mode via collective light scattering at rate $\lambda$. We can then expand the atomic field operators in terms of a two level basis of momentum states $\psi(x,z)=c_0 \psi_0(x,z) + c_1 \psi_1(x,z)$. When we come to the description of the many-body system in the second quantisation formalism, the coefficients of the decomposition are bosonic operators $\hat{c}_0$ and $\hat{c}_1$, which are mapped into collective angular momentum operators via the Schwinger transformation
\begin{equation}
\hat{J}_-=\hat{c}_0^{\dagger}\hat{c}_1\hspace{0.5cm}\hat{J}_+=\hat{J}_-^{\dagger}\hspace{0.5cm}\hat{J_z}=\left(\hat{c}_1^{\dagger}\hat{c}_1-\hat{c}_0^{\dagger}\hat{c}_0\right)/2.
\end{equation}
This procedure realises a mapping to the effective Hamiltonian (with $\hbar=1$)
\begin{equation}
\hat{H}=\omega_0 \hat{J}_z+\omega \hat{a}^{\dagger}\hat{a}+\frac{2\lambda}{\sqrt{N}}\left(\hat{a}+\hat{a}^{\dagger}\right)\hat{J}_x,
\end{equation}
where $\omega$ is the detuning between the cavity mode frequency and the transverse pump frequency, and $\lambda$ is proportional to the square root of the intensity of the pump. The conservation of the atom number, $\hat{c}_1^{\dagger}\hat{c}_1+\hat{c}^{\dagger}_0\hat{c}_0=N$, in this case directly implies the conservation of the Dicke cooperation number $J=N/2$.\\
The open nature of the system is due to photons escaping the cavity through a loss channel at rate $\kappa$, and the density fluctuations can then be inferred from the detected cavity output field. Furthermore we keep into account an atomic dissipation channel in a phenomenological way  by including a damping rate $\gamma_b$ for the atomic motional degree of freedom. This dissipation channel is due to collisional or cavity-mediated coupling of excitations of the excited momentum mode to Bogoliubov modes with wave vectors that are different  from that of the pump and cavity fields \cite{SdickeETH}. The collection of these modes provides a heat-bath at the condensate temperature $T$, that we assume to be of Markovian nature.

\subsection{Dicke Model with cavity dissipation}
The Dicke Hamiltonian was originally introduced to describe the coupling between an ensemble of $N$ two-level atoms and a single cavity mode \cite{SDicke}. We will consider the Hamiltonian of the Dicke model in the form
\begin{equation}
\hat{H}=\omega_0 \hat{J}_z+\omega \hat{a}^{\dagger}\hat{a}+\frac{2\lambda}{\sqrt{N}}\left(\hat{a}+\hat{a}^{\dagger}\right)(\hat{J}_x+\zeta)
\end{equation}
where $\zeta \in {\cal R}$ denotes an explicit symmetry breaking field. We have defined collective atomic angular momentum operators $\hat{J}_{\alpha}$ $(\alpha=x,y,z)$ and bosonic field mode operators $\hat{a}$ and $\hat{a}^{\dagger}$.
We can define the mean fields
\begin{equation}
\langle \hat{a}\rangle=\alpha, \hspace{1cm} \langle \hat{J}_- \rangle=\beta, \hspace{1cm} \langle \hat{J}_z \rangle=w
\end{equation}
and write the semiclassical equations of motion including a cavity decay at rate $\kappa$
\begin{equation}
\begin{aligned}
\dot{\alpha}&=-(\kappa+i \omega)\alpha-i\frac{\lambda}{\sqrt{N}}\left(\beta+\beta^*+2\zeta \right),\\
\dot{\beta}&=-i \omega_0 \beta+2i \frac{\lambda}{\sqrt{N}}\left(\alpha+\alpha^*\right)w,\\
\dot{w}&=i\frac{\lambda}{\sqrt{N}}(\alpha+\alpha^*)(\beta-\beta^*).\\
\end{aligned}
\end{equation}
Using the angular momentum conservation $w^2+|\beta|^2=N^2/4$, we get the steady-state equations
\begin{equation}
\begin{aligned}
\label{meanfields}
\beta_{s}&=\left(\frac{\lambda}{\lambda_{\text{cr}}}\right)^2 \left(\beta_{s}+\zeta\right)\sqrt{1-4\frac{\beta_{s}^2}{N^2}},\\
\alpha_{s}&=\frac{2\lambda}{i\kappa -\omega}\frac{\left(\beta_{s}+\zeta\right)}{\sqrt{N}},\\
\end{aligned}
\end{equation}
where the critical coupling strength is $\lambda_{\text{cr}}=\frac{1}{2}\sqrt{\frac{\omega_0}{\omega}\left(\kappa^2+\omega^2\right)}$ for $\zeta=0$. Notice that $\beta_{s}$ is real and $\beta_{s} = O(1/N)$, while $\alpha_{s}$ is complex and $\alpha_{s}= O(1/\sqrt{N})$. Now we rewrite the Hamiltonian explicitly in terms of the operators which represent displacements of atomic and field operators with respect to the stationary values of the respective mean fields $\beta_{s}$ and $\alpha_{s}$. It is convenient to apply the Holstein-Primakoff transformation:
\begin{equation}
\label{schw}
\hat{J}_+=\hat{b}^{\dagger} \sqrt{N-\hat{b}^{\dagger}\hat{b}}, \hspace{0.2cm} \hat{J}_-=\sqrt{N-\hat{b}^{\dagger}\hat{b}}\hspace{0.1cm} \hat{b},\hspace{0.2cm} \hat{J}_z =\hat{b}^{\dagger}\hat{b}-\frac{N}{2},
\end{equation}
and then introduce the fluctuations operators:
\begin{equation}
\delta\hat{a}=\hat{a}-\widetilde{\alpha}, \hspace{1cm} \delta \hat{b}=\hat{b}-\frac{\widetilde{\beta}}{\sqrt{N}},
\end{equation}
where $\widetilde{\alpha}$ and $\widetilde{\beta}$ are the steady-state mean fields of the operators $\hat{a}$ and $\hat{b}$ respectively. We easily recognise $\widetilde{\alpha}=\alpha_{s}$. The relation between the steady-state mean value of $\hat{J}_-$ $(\beta_s)$ and the mean value of the bosonic operator $\hat{b}$ $(\widetilde{\beta}/\sqrt{N})$ is given instead by an expansion at the thermodynamic limit ($N\gg1$) of the mean value of $\hat{J}_-$ as given by Eq.\eqref{schw}. The final relation is given by
\begin{equation}
\label{betass}
\widetilde{\beta}\sqrt{1-\frac{\widetilde{\beta}^2}{N^2}}=\beta_{s}.
\end{equation}
With these definitions the leading term for the time evolution of the density matrix of the total system in the thermodynamical limit is given by:
\begin{equation}
\frac{d \hat{\rho}}{dt}=-i \left[\hat{H}',\hat{\rho}\right]+{\cal L}'(\hat{\rho})
\end{equation}
where the Hamiltonian, neglecting constant terms, is given by
\begin{equation}
\begin{aligned}
\label{DickeDisplaced}
\hat{H}'&=\tilde{\omega}_0 \delta\hat{b}^{\dagger}\delta\hat{b}+\omega \delta\hat{a}^{\dagger}\delta\hat{a}+\widetilde{\lambda}(\delta\hat{a}+\delta\hat{a}^{\dagger})(\delta\hat{b}+\delta\hat{b}^{\dagger})\\
&-\mu \left(\delta\hat{b}+\delta\hat{b}^{\dagger}\right)^2
\end{aligned}
\end{equation}
with parameters
\begin{equation}
\begin{aligned}
\label{parameterss}
\tilde{\omega}_0&=\omega_0-\frac{2\lambda\Re{({\widetilde{\alpha}})}{\beta}}{N^{3/2}\sqrt{1-\frac{\widetilde{\beta}^2}{N^2}}},\\
\mu&=\frac{\lambda \Re{({\widetilde{\alpha}})}\widetilde{\beta}}{N^{3/2}\sqrt{1-\frac{\widetilde{\beta}^2}{N^2}}}\left(1+\frac{\widetilde{\beta}^2}{2(N^2-\widetilde{\beta}^2)}\right),\\
\widetilde{\lambda}&=\lambda \frac{1-2\frac{\widetilde{\beta}^2}{N^2}}{\sqrt{1-\frac{\widetilde{\beta}^2}{N^2}}}.\\
\end{aligned}
\end{equation}
The non-unitary evolution, with only cavity dissipation present, is given by the Linblad superoperator
\begin{equation}
\label{linblad}
{\cal L}'\left(\hat{\rho}\right)=\kappa \left(2\delta\hat{a}\hspace{0.1cm}\hat{\rho}\hspace{0.1cm}\delta\hat{a}^{\dagger}-\delta\hat{a}^{\dagger}\delta\hat{a}\hat{\rho}-\hat{\rho}\delta\hat{a}^{\dagger}\delta\hat{a}\right).\\
\end{equation}
Considering also the atomic dissipation channel, the semiclassical steady-state is not significantly influenced by the weak atomic dissipation rate $\gamma_b$. Thus we can describe the system with the hamiltonian $\hat{H}'$ in Eq.~\eqref{DickeDisplaced}, displaced by the semiclassical steady-state amplitudes $\widetilde{\alpha}$ and $\widetilde{\beta}$ in Eqs.~\eqref{meanfields} and \eqref{betass} due to only the cavity dissipation channel, while considering to total dissipation Linblad superoperator
\begin{equation}
\begin{aligned}
\label{linblad2}
{\cal L}''\left(\hat{\rho}\right)&=\kappa \left(2\delta\hat{a}\hspace{0.1cm}\hat{\rho}\hspace{0.1cm}\delta\hat{a}^{\dagger}-\delta\hat{a}^{\dagger}\delta\hat{a}\hat{\rho}-\hat{\rho}\delta\hat{a}^{\dagger}\delta\hat{a}\right)+\\
&+\gamma_b(n_T+1) \left(2\delta\hat{b}\hspace{0.1cm}\hat{\rho}\hspace{0.1cm}\delta\hat{b}^{\dagger}-\delta\hat{b}^{\dagger}\delta\hat{b}\hat{\rho}-\hat{\rho}\delta\hat{b}^{\dagger}\delta\hat{b}\right)+\\
&+\gamma_b n_T \left(2\delta\hat{b}^{\dagger}\hspace{0.1cm}\hat{\rho}\hspace{0.1cm}\delta\hat{b}-\delta\hat{b}\delta\hat{b}^{\dagger}\hat{\rho}-\hat{\rho}\delta\hat{b}\delta\hat{b}^{\dagger}\right).\\
\end{aligned}
\end{equation}
In the above expression $n_T=(\exp(\hbar \tilde{\omega}_0/k_b T)-1)^{-1}$ is the average number of thermal atom excitations at the condensate temperature $T$.
The correspondence between the notation used in this section for the description of the cavity-BEC system and the common one used in the main text is summarised in the following table
\begin{center}
\begin{tabular}{ |c||c|c|c| }
\hline
 & Cavity-BEC notation & Our common notation \\
\hline
\hline
\multirow{3}{4em}{Atomic mode} & $\hat{q},\, \hat{p}$ & $\hat{q}_b,\, \hat{p}_b$ \\ 
& $\tilde{\omega}_0$ & $\omega_b$ \\ 
& $\gamma_b, \, n_T$ & $\gamma_b, \, n_{T_b}$ \\ 
\hline
\multirow{3}{4em}{Cavity field} & $\hat{x},\, \hat{y}$ & $\hat{q}_a,\, \hat{p}_a$ \\ 
& $\omega$ & $\omega_a$ \\ 
& $\kappa$ & $\kappa_a$ \\
\hline
\multirow{1}{4em}{Coupling} & $\widetilde{\lambda}$ & $g_{ab}$ \\
\hline
\end{tabular}
\end{center}

\subsection{Quantum Langevin Equations}

The quadratic Hamiltonian \eqref{DickeDisplaced} and the Linblad superoperator \eqref{linblad2} are responsible for linear equations of motion for the operators $\delta\hat{a}$ and $\delta\hat{b}$, that in the equivalent formalism of quantum Langevin equations are given by
\begin{equation}
\label{openAtoms}
\begin{aligned}
\delta \dot{\hat{a}}&=i\left[\hat{H'},\delta\hat{a}\right]-\kappa\delta\hat{a}+\sqrt{2\kappa}\hat{a}^{\text{in}}\\
\delta\dot{\hat{b}}&=i\left[\hat{H'},\delta\hat{b}\right]-\gamma_b\delta\hat{b}+\sqrt{2\gamma_b}\hat{b}^{\text{in}}\\
\end{aligned}
\end{equation}
and relative hermitian conjugate equations. The operators $\hat{a}^{\text{in}}$ and $\hat{b}^{\text{in}}$ are the input noise operators, i.e. bath operators responsible for additional fluctuations, with respect to the closed case, of the system observables. The time correlation relations for the input noise operators are specified by the thermal nature of the baths. In our case the continuum of modes of the electromagnetic field outside the cavity constitutes a bath at zero excitations.
%since no thermal photons are excited at room temperature in the optical spectral range.
The atomic heat-bath instead is at the condensate temperature $T$ \cite{SdickeETH}. The time correlation functions for the noise operators are
\begin{equation}
\begin{aligned}
&\langle\hat{a}^{\text{in}}(t)\hat{a}^{\text{in} \dagger}(t')\rangle=\delta(t-t'),\hspace{0.5cm}\langle\hat{a}^{\text{in}\dagger}(t)\hat{a}^{\text{in}}(t')\rangle=0,\\
&\langle\hat{b}^{\text{in}}(t)\hat{b}^{\text{in}\dagger}(t')\rangle=(n_T+1)\delta(t-t'),\\
&\langle\hat{b}^{\text{in}\dagger}(t)\hat{b}^{\text{in}}(t')\rangle=n_T\delta(t-t').\\
\end{aligned}
\end{equation}
If we now define the quadrature operators for both the atomic and field systems
\begin{equation}
\begin{aligned}
\delta\hat{x}&=\left(\delta\hat{a}+\delta\hat{a}^{\dagger}\right)/\sqrt{2} \hspace{1cm}\delta\hat{y}&=i\left(\delta\hat{a}^{\dagger}-\delta\hat{a}\right)/\sqrt{2}\\
\delta\hat{q}&=\left(\delta\hat{b}+\delta\hat{b}^{\dagger}\right)/\sqrt{2} \hspace{1cm }\delta\hat{p}&=i\left(\delta\hat{b}^{\dagger}-\delta\hat{b}\right)/\sqrt{2},\\
\end{aligned}
\end{equation}
the vectorial quantum Lagevin equation assumes the form
\begin{equation}
\dot{u}(t)=Au+N(t),
\end{equation}
where $u(t) = (\delta \hat{q}, \delta \hat{p}, \delta \hat{x}, \delta \hat{y})$, 

\[
A =
\begin{pmatrix}
-\gamma_b & \tilde{\omega}_0& 0 & 0 \\
-(\tilde{\omega}_0-4\mu) & -\gamma_b& -2\widetilde{\lambda} & 0 \\
0 & 0& -\kappa & \omega \\
-2 \widetilde{\lambda} & 0& -\omega & -\kappa \\
\end{pmatrix},
\]
and $N(t)=(\sqrt{2\gamma_b}\hat{q}^{\text{in}},\sqrt{2\gamma_b}\hat{p}^{\text{in}},\sqrt{2\kappa}\hat{x}^{\text{in}},\sqrt{2\kappa}\hat{y}^{\text{in}})$.
The covariance matrix $\sigma$ satisfies the equation
\begin{equation}\label{sigma_dif}
\frac{\ud \sigma}{\ud t} = A \sigma + \sigma A\trans + D,
\end{equation}
with $D=\text{diag}\left(\gamma_b(2 n_T+1),\gamma_b(2 n_T+1),\kappa,\kappa\right)$. 
The covariance matrix $\sigma_s$ of the non-equilibrium steady-state is a solution of the Lyapunov equation $A \sigma_s + \sigma_s A\trans = - D$.

\subsection{Polariton Modes}
The coupling between the atomic and photonic degrees of freedom gives rise to polariton modes which can be defined via diagonalization of the two oscillators system. Our definition of polariton modes will be based on the diagonalization of the closed system similarly to Ref.\cite{SEmaryNew}. Diagonalization of the closed system is accomplished via definition of the symplectic matrix
\begin{equation}\label{M}
M = \begin{pmatrix}
A&B&G&D\\
B&A&D&G\\
A_2&B_2&G_2&D_2\\
B_2&A_2&D_2&G_2\\
\end{pmatrix}
\end{equation}
and vectors $\mathbf{a}=(\delta\hat{a},\delta\hat{a}^{\dagger},\delta\hat{b},\delta\hat{b}^{\dagger})^T$ and $\mathbf{d}=(\hat{d},\hat{d}^{\dagger},\hat{c},\hat{c}^{\dagger})^T$, so that we can write the matrix equation $\mathbf{a}=M \mathbf{d}$. The parameters defining the diagonalization are given in Appendix.
In terms of the polariton operators $\mathbf{d}=(\hat{d},\hat{d}^{\dagger},\hat{c},\hat{c}^{\dagger})^T$ the Hamiltonian will assume the diagonal form
\begin{equation}
\label{polaritondiagonal}
\hat{H'}=\epsilon_- \hat{d}^{\dagger}\hat{d}+\epsilon_+ \hat{c}^{\dagger}\hat{c},
\end{equation}
neglecting constant terms. The eigenvalues are given by
\begin{widetext}
\begin{equation}
\label{energies}
\epsilon_{\pm}=\sqrt{\frac{1}{2}\left\{\omega^2+\tilde{\omega}_0^2-4\mu \tilde{\omega}_0\pm\text{sign}\left(\tilde{\omega}_0^2-\omega^2-4 \mu\tilde{\omega}_0\right)\sqrt{\left(\tilde{\omega}_0^2-\omega^2-4 \mu\tilde{\omega}_0\right)^2+16\widetilde{\lambda}^2\omega\tilde{\omega}_0}\right\}}
\end{equation}
\end{widetext}
where $\epsilon_+$ is the softening frequency. In the regime of parameters of the experiment performed (See Table in the main text) the softening frequency can be approximated by
\begin{equation}
\label{omeganormal}
\frac{\epsilon_+}{\omega_0}\approx\sqrt{1-x\left[1+\left(\frac{\kappa}{\omega}\right)^2\right]}\approx \sqrt{1-x}
\end{equation}
for $x=\left(\lambda/\lambda_{\text{cr}}\right)^2<1$ (normal phase), and
\begin{equation}
\label{omegasuper}
\frac{\epsilon_+}{\omega_0}\approx\sqrt{x^2-1-\left(\frac{\kappa}{\omega}\right)^2}\approx \sqrt{x^2-1}
\end{equation}
for $x=\left(\lambda/\lambda_{\text{cr}}\right)^2>1$ (superradiant phase).\\ We notice that the critical exponent for the scaling of the soft frequency is the same in the two phases with a prefactor of $\sqrt{2}$ for the super phase.

\subsection{Master Equation}
The description of the open dynamics of the system is based on the hierarchy of the parameters that determine the time scales \cite{SZurichNew}. As the cavity decay rate $\kappa$ is the fastest rate, we solve first the master equation with only the cavity dissipation. After diagonalization, we then introduce the dissipation channel for the atomic polariton mode $\hat{c}$ at a rate $\gamma_c$. This is due to the fact that the atomic damping happens in a long time scale, given by the small decay rate $\gamma_b$, after the atoms have been dressed with photons of the cavity mode on the relevant time scale given by $\lambda$. As stated above the time evolution of the system with only cavity dissipation is described by the master equation
\begin{equation}
\frac{d \hat{\rho}}{dt}=-i \left[\hat{H}',\hat{\rho}\right]+{\cal L}'(\hat{\rho}),
\end{equation}
where
\begin{equation}
{\cal L}'\left(\hat{\rho}\right)=\kappa \left(2\delta\hat{a}\hspace{0.1cm}\hat{\rho}\hspace{0.1cm}\delta\hat{a}^{\dagger}-\delta\hat{a}^{\dagger}\delta\hat{a}\hat{\rho}-\hat{\rho}\delta\hat{a}^{\dagger}\delta\hat{a}\right).\\
\end{equation}
The diagonalization mixes all four operators in the following way
\begin{equation}
\begin{aligned}
\label{diagon}
\delta\hat{a}&=A \hat{d}+B \hat{d}^{\dagger}+G \hat{c}+D \hat{c}^{\dagger}\\
\delta\hat{a}^{\dagger}&=B \hat{d}+A \hat{d}^{\dagger}+D \hat{c}+G \hat{c}^{\dagger},
\end{aligned}
\end{equation}
and similar expression for $\delta \hat{b}$ and $\delta \hat{b}^{\dagger}$.
While this operation effectively diagonalises the Hamiltonian in the form \eqref{polaritondiagonal},  it also makes the non unitary term of the master equation much more involved giving rise in principle to 16 different terms. However, in the interaction picture with respect to the free polariton Hamiltonian, we can perform a rotating wave approximation that allows to obtain the simpler master equation
\begin{equation}
\label{masterpolar}
\begin{aligned}
\frac{d \hat{\rho}}{dt}&=\kappa A^2\left(2 \hat{d}\hat{\rho}\hat{d}^{\dagger}-\hat{d}^{\dagger}\hat{d}\hat{\rho}-\hat{\rho}\hat{d}^{\dagger}\hat{d}\right)+\\
+&\kappa B^2\left(2 \hat{d}^{\dagger}\hat{\rho}\hat{d}-\hat{d}\hat{d}^{\dagger}\hat{\rho}-\hat{\rho}\hat{d}\hat{d}^{\dagger}\right)+\\
+&\left(\kappa G^2+\gamma_c(n_{T}^c+1)\right)\left(2 \hat{c}\hat{\rho}\hat{c}^{\dagger}-\hat{c}^{\dagger}\hat{c}\hat{\rho}-\hat{\rho}\hat{c}^{\dagger}\hat{c}\right)+\\
+&\left(\kappa D^2+\gamma_c n_T^c\right)\left(2 \hat{c}^{\dagger}\hat{\rho}\hat{c}-\hat{d}\hat{c}^{\dagger}\hat{\rho}-\hat{\rho}\hat{c}\hat{c}^{\dagger}\right).
\end{aligned}
\end{equation}
The validity of this approximation is justified by the agreement with the experimental data. In the equation above we have introduced the atomic polariton dissipation channel with rate $\gamma_c$. The parameter $n_T^c=(\exp(\hbar \omega_S/k_B T)-1)^{-1}$ gives the thermal average of occupation of the atomic polariton mode $\hat{c}$, as given by the Bose-Einstein distribution function evaluated at the soft mode frequency $\omega_S\equiv\epsilon_+$.
%The master equation \eqref{masterpolar}describes two independent oscillators coupled to two independent thermal baths.
A comparison with an ab initio treatment of the open dynamics with atomic losses at rate $\gamma_b$, as defined in Eq.~\eqref{openAtoms}, makes possible to identify the effective atomic damping rate in terms of the atomic polariton damping rate, as given by the equation
\begin{equation}
\label{gammaB}
\gamma_b=\frac{\gamma_c(n_T^c+1)}{(n_T+1)G_2^2+n_T D_2^2}.
\end{equation}
\begin{figure}[h!]
\centering
\includegraphics[width=0.5\textwidth]{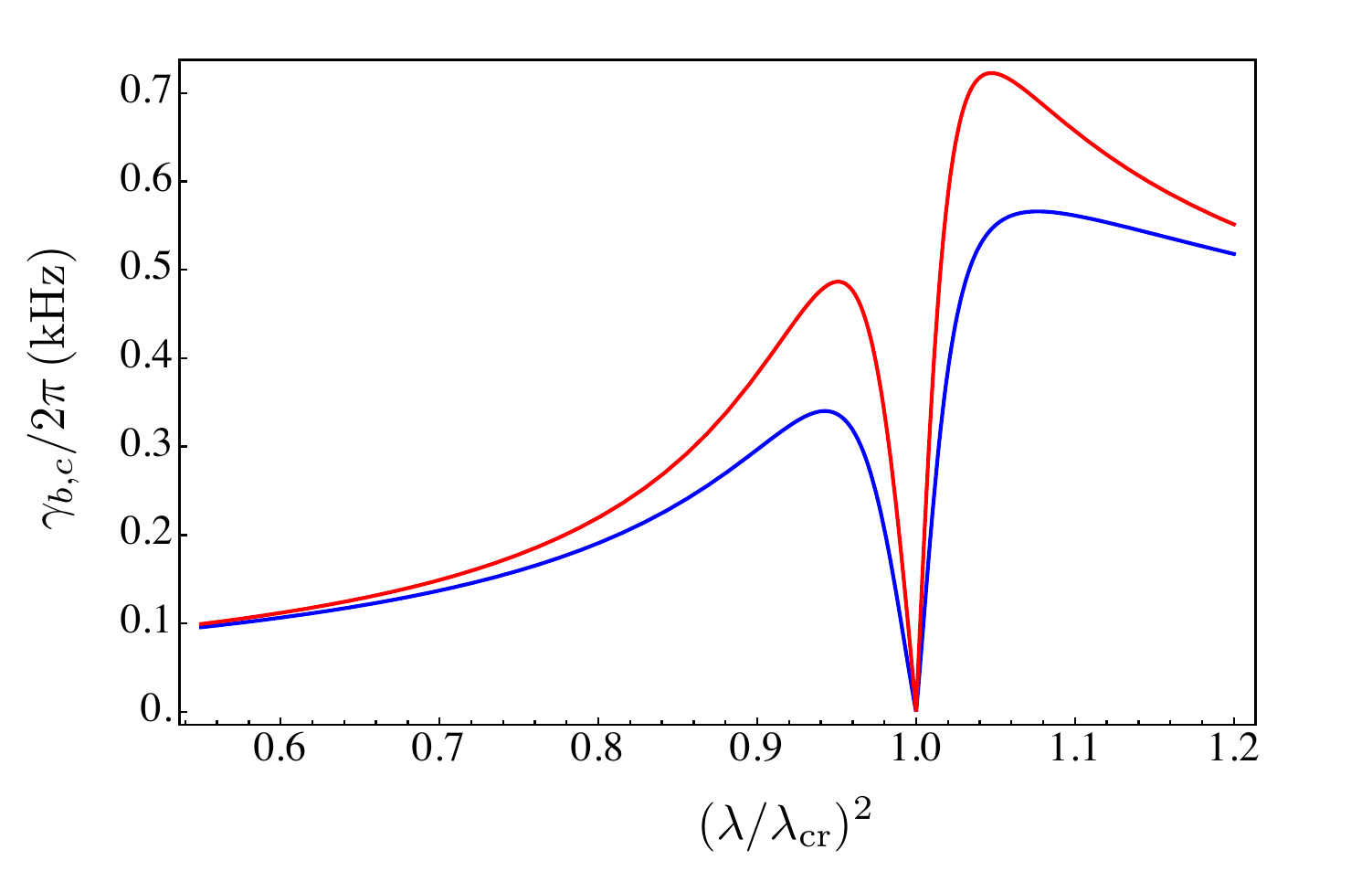} 
\caption{Red curve: atomic polariton damping rate $\gamma_c$, as obtained from the fit of the experimental values. Blue curve: effective atomic damping rate $\gamma_b$, as defined in Eq.~\eqref{gammaB}.}
\label{gammas}
\end{figure}
Fig.\ref{gammas} clearly shows that the atomic polariton damping rate $\gamma_c$ is bigger than the bare atomic damping rate $\gamma_b$. This is physically due to the fact the the small population of photons dressing the atoms, which are characterised by the fast damping rate $\kappa$, contribute to redefine an effective damping rate for the atomic polariton mode $\gamma_c$ bigger than the bare atomic one. Also we see that for small couplings, where the atomic polariton mode $\hat{c}$ is not much different than the bare atomic mode $\delta\hat{b}$, the two damping rates are very similar.

\subsection{Measurement}

The occupation $\langle \hat{c}^{\dagger} \hat{c}\rangle$ of the quasi-particle mode can be extracted from the observed sideband asymmetry. The experimental data set for the excitations of quasi-particles  $\langle \hat{c}^{\dagger}\hat{c}\rangle_{\text{exp}}$ is obtained summing $n_T^c$, evaluated at the experimental values of $\omega_S$, to the experimental set of $\langle\hat{c}^{\dagger}\hat{c}\rangle-n_T^c$. The latter is measured via integration over the sidebands of the density noise spectrum of light, by making use of the rate equation
\begin{equation}
\label{rateequation}
2 \kappa \left(\langle \delta \hat{a}^{\dagger}\delta \hat{a}\rangle_- - \langle \delta \hat{a}^{\dagger}\delta \hat{a}\rangle_+\right)=2 \gamma \left(\langle \hat{c}^{\dagger}\hat{c}\rangle-n_T\right),
\end{equation}
valid at the steady-state~\cite{SZurichNew}. In this equation $\langle \delta \hat{a}^{\dagger}\delta \hat{a}\rangle_\pm$ is the integrated spectral weight of the blue and red sideband, respectively. The rate equation above expresses a balance between  the total rate at which quasi-particles are created $2 \kappa \langle \delta \hat{a}^{\dagger}\delta \hat{a}\rangle_- +2 \gamma n_T$, and the total rate at which quasi-particles are annihilated $2 \kappa \langle \delta \hat{a}^{\dagger}\delta \hat{a}\rangle_+ +2 \gamma \langle \hat{c}^{\dagger}\hat{c} \rangle$.\\A balanced heterodyne scheme allows us to measure the density noise spectrum of light outside the cavity. With a standard input-output theory we are then able to infer, excluding the coherent part, the separate integrated sidebands of the spectrum of light inside the cavity, and so the average number of occupation of the quasi-particles mode $\hat{c}$, as given by Eq.~\eqref{rateequation}.\\
Keeping in mind the transformation \eqref{diagon} (and similar for $\delta\hat{b}$ and $\delta\hat{b}^{\dagger}$) we notice that both the average number of fluctuations of photons $n_a=\langle \delta \hat{a}^{\dagger}\delta \hat{a}\rangle$ and atomic occupation $n_b=\langle \delta \hat{b}^{\dagger}\delta \hat{b}\rangle$ can be expressed in terms of $\langle \hat{c}^{\dagger}\hat{c}\rangle$. Thus we can obtain two sets of experimental values for $n_a$ and $n_b$ via the relation
\begin{equation}
\begin{aligned}
\label{experimental}
(n_a)_{\text{exp}}&=B^2+D^2+\left(D^2+G^2\right)\langle \hat{c}^{\dagger} \hat{c}\rangle_{\text{exp}}\\
(n_b)_{\text{exp}}&=B_2^2+D_2^2+\left(D_2^2+G_2^2\right)\langle \hat{c}^{\dagger} \hat{c}\rangle_{\text{exp}}.\\
\end{aligned}
\end{equation} 
where we have used the condition that the occupation of the photonic polariton mode $\langle \hat{d}^{\dagger}\hat{d}\rangle$ is vanishingly small.\\
Thus measuring $\gamma_c$ and $\langle \hat{c}^{\dagger}\hat{c}\rangle$, we are able to reconstruct, with Eqs.~\eqref{gammaB} and \eqref{experimental}, the effective atomic damping rate $\gamma_b$ and the average occupations of the \emph{local} modes $\delta\hat{a}$ and $\delta\hat{b}$, and so the irreversible entropy production rate at the steady-state $\Pi_s$, as given by Eq.~\eqref{Pi_ss}.

\section{Appendix}
\subsection{Diagonalization parameters}
\begin{equation*}
\begin{aligned}
A&=\frac{1}{2}\cos \left(\gamma_S\right)\left(\sqrt{\frac{\omega}{\epsilon_-}}+\sqrt{\frac{\epsilon_-}{\omega}}\right),\\
B&=\frac{1}{2}\cos \left(\gamma_S\right)\left(\sqrt{\frac{\omega}{\epsilon_-}}-\sqrt{\frac{\epsilon_-}{\omega}}\right),\\
G&=\frac{1}{2}\sin \left(\gamma_S\right)\left(\sqrt{\frac{\omega}{\epsilon_+}}+\sqrt{\frac{\epsilon_+}{\omega}}\right),\\
D&=\frac{1}{2}\sin \left(\gamma_S\right)\left(\sqrt{\frac{\omega}{\epsilon_+}}-\sqrt{\frac{\epsilon_+}{\omega}}\right),\\
\end{aligned}
\end{equation*}

\begin{equation*}
A_2=-\frac{1}{2}\sin \left(\gamma_S\right)\left(\sqrt{\frac{\tilde{\omega}_0}{\epsilon_-}}+\sqrt{\frac{\epsilon_-}{\tilde{\omega}_0}}\right),
\end{equation*}

\begin{equation*}
B_2=-\frac{1}{2}\sin \left(\gamma_S\right)\left(\sqrt{\frac{\tilde{\omega}_0}{\epsilon_-}}-\sqrt{\frac{\epsilon_-}{\tilde{\omega}_0}}\right),
\end{equation*}
\begin{equation*}
G_2=\frac{1}{2}\cos \left(\gamma_S\right)\left(\sqrt{\frac{\tilde{\omega}_0}{\epsilon_+}}+\sqrt{\frac{\epsilon_+}{\tilde{\omega}_0}}\right),
\end{equation*}

\begin{equation*}
D_2=\frac{1}{2}\cos \left(\gamma_S\right)\left(\sqrt{\frac{\tilde{\omega}_0}{\epsilon_+}}-\sqrt{\frac{\epsilon_+}{\tilde{\omega}_0}}\right),
\end{equation*}
with the Bogoliubov angle
\begin{equation}
\tan \left(2 \gamma_S \right)=\frac{4 \widetilde{\lambda}\sqrt{\omega \tilde{\omega}_0}}{\tilde{\omega}_0^2-4\mu \tilde{\omega}_0-\omega^2}.\\
\end{equation}

%\clearpage


\begin{thebibliography}{99}
\bibitem{footnote}
The observed atomic damping, for the cavity-BEC setup, takes place on time scales longer than the time scale of the appearance of polariton modes. Thus the Bose distribution function, for the thermal number of excitation, is evaluated at the softening frequency of the atomic polariton mode (see SI).
\end{thebibliography}
\end{document}